\documentstyle[12pt,amsfonts]{article}
\parskip=\medskipamount

\def\appendix#1{
  \addtocounter{section}{1}
  \setcounter{equation}{0}
  \renewcommand{\thesection}{\Alph{section}}
 \section*{Appendix \thesection\protect\indent
 \parbox[t]{11.715cm} {#1}} 
 \addcontentsline{toc}{section}{Appendix \thesection\ \ \ #1}
  }

\renewcommand{\thefootnote}{\fnsymbol{footnote}}

\newcommand {\cD}{{\cal D}}

\newcommand {\cF}{{\cal F}}

\newcommand {\cL}{{\cal L}}
\newcommand {\cM}{{\cal M}}
\newcommand {\cN}{{\cal N}}
\newcommand {\cO}{{\cal O}}
\newcommand {\cP}{{\cal P}}
\newcommand {\cQ}{{\cal Q}}

\newcommand {\cS}{{\cal S}}

\newcommand {\cW}{{\cal W}}
\newcommand {\cX}{{\cal X}}
\newcommand {\cY}{{\cal Y}}
\newcommand {\cZ}{{\cal Z}}
\def\a{\alpha}
\def \bi{\bibitem}

\def\b{\beta}

\def\d{\delta}
\def\e{\epsilon}
\def\f{\phi}
\def\g{\gamma}
\def\G{\Gamma}

\def\j{\psi}
\def\k{\kappa}
\def\l{\lambda}
\def\m{\mu}
\def\n{\nu}
\def\o{\omega}
\def\p{\pi}
\def\q{\theta}

\def\s{\sigma}

\def\x{\xi}

\def\D{\Delta}
\def\F{\Phi}
\def\J{\Psi}
\def\L{\Lambda}
\def\O{\Omega}

\def\U{\Upsilon}

\newcommand{\ad}{{\dot{\alpha}}}                           
\newcommand{\bd}{{\dot{\beta}}}                            
\newcommand{\ve}{\varepsilon}                            

\newcommand{\pa}{\partial}                           
\newcommand{\hf}{\frac12}

\newcommand{\vf}{\varphi}
\newcommand{\sect}[1]{\setcounter{equation}{0}\section{#1}}

\newcommand{\be}{\begin{equation}}
\newcommand{\ee}{\end{equation}}
\newcommand{\bea}{\begin{eqnarray}}
\newcommand{\eea}{\end{eqnarray}}
\newcommand{\non}{\nonumber}
\newcommand{\1}{\underline{1}}
\newcommand{\2}{\underline{2}}
\begin{document}

\begin{titlepage}
\thispagestyle{empty}

\begin{flushright}
LMU-TPW-00-19\\
hep-th/0007231 \\
July, 2000
\end{flushright}
\vspace{5mm}

\begin{center}
{\Large\bf  Nonlinear Self--Duality and Supersymmetry}\footnote{Based on talks
given at the XII Workshop `Beyond the Standard Model'
(February 2000, Bad Honnef, Germany) and at the Erwin Schr\"odinger International 
Institute for Mathematical Physics (March 2000, Vienna, Austria).} 
\end{center}

\begin{center} 
{\large 
Sergei M. Kuzenko\footnote{Address after September 1, 2000:
Department of Physics, The University of Western Australia, 
Nedlands, W.A. 6907, Australia.} 
 and Stefan Theisen\footnote{Address after August 1, 2000:
Max-Planck Institut f\"ur Gravitationsphysik,
Albert-Einstein-Institut, Am M\"uhlenberg 1, 
D-14476 Golm, Germany.}
}\\
\vspace{2mm}

${}$\footnotesize{
{\it Sektion Physik, Universit\"at M\"unchen\\
Theresienstr. 37, D-80333 M\"unchen, Germany} 
} \\
{\tt sergei@theorie.physik.uni-muenchen.de} \\
{\tt theisen@theorie.physik.uni-muenchen.de}
\vspace{2mm}

\end{center}
\vspace{5mm}
                  
\begin{abstract}
\baselineskip=14pt
We review self-duality of nonlinear electrodynamics and its 
extension to several Abelian gauge fields coupled to 
scalars. We then describe self-duality in 
supersymmetric models, both ${\cal N}=1$ and ${\cal N}=2$.
The self-duality equations, which have to be satisfied by 
the action of any self-dual system, are found and 
solutions are discussed. One important example is the 
Born-Infeld action. We explain why the ${\cal N}=2$ 
supersymmetric actions proposed so far are not the 
correct world-volume actions for D3 branes in $d=6$. 
\end{abstract}
\vfill
\end{titlepage}

\newpage
\setcounter{page}{1}

\renewcommand{\thefootnote}{\arabic{footnote}}
\setcounter{footnote}{0}
\sect{Introduction}
The simplest and best known example of a self-dual system is 
electrodynamics in vacuum. 
The set of Maxwell's equations is invariant under the 
simultaneous replacements $\vec E\to \vec B,\,\vec B\to -\vec E$. 
While being a symmetry of the Hamiltonian $H=\vec E^2+\vec B^2$, 
the Lagrangian does transform: $L=\vec E^2-\vec B^2\to -L$. 
The generalization to a $(p-1)$-form potential $C$ in 
$d=2p$ dimensions with action $S=\int {\rm d}C\wedge * {\rm d}C$ 
is immediate. 

These theories are free systems with linear equations of motion. 
The interesting question is whether one can construct 
interacting self-dual systems. 
The main goal of these notes is to 
discuss the conditions ({\it self-duality equations}) which have 
to be satisfied by the action of a dynamical system in order to be 
self-dual, in the sense to be specified below.
Apparently Schr\"odinger was the first to discuss
nonlinear self-duality. In \cite{Sch} he reformulated
the Born-Infeld (BI) theory \cite{BI} in such a way 
that it was manifestly invariant under
U(1) duality rotations.
We will mainly be interested in four-dimensional 
nonlinear systems of gauge fields  coupled to matter. 
For non-supersymmetric systems the results have been 
obtained, as a generalization of patterns of duality in 
extended supergravity \cite{FSZ,CJ} (see also \cite{CJLP}), 
in \cite{GZ1,Z,GR1,GR2,GZ2,GZ3} and reviewed and 
extended in \cite{Ta,AT}. Our special emphasis is on manifestly 
${\cal N}=1,2$ supersymmetric generalizations. 

As will be discussed below, self-dual theories
possess quite remarkable properties.
Our main concern, however, in pursuing the study of
such systems lies in the fact that self-duality
turns out to be intimately connected with spontaneous 
breaking of supersymmetry (for still not completely understood 
reasons). Recently several models for partial breaking of
$\cN=2$ supersymmetry to $\cN=1$ in four dimensions
\cite{BG,BG2,RT,GPR} have been constructed. 
Two most prominent models -- 
described by the Goldstone-Maxwell multiplet \cite{BG,RT}
and by the tensor Goldstone multiplet \cite{BG2,RT} --
are self-dual $\cN=1$ supersymmetric theories; the other 
Goldstone multiplets are dual superfield version of the 
tensor one (as we will describe, 
self-duality may be consistent with the existence 
of dual formulations). In our opinion, this cannot be 
accidental.

It may look curious but the fact that the nonlinear superfield
constraint, which underlies the Goldstone-Maxwell construction 
of \cite{BG,RT}, has turned out to be fruitful for 
nontrivial generalizations. This constraint was used in 
\cite{BMZ,ABMZ} to derive nonlinear
U$(n)$ duality invariant models, both in non-supersymmetric
and supersymmetric cases. In the present paper, 
we apply the nonlinear constraint, which is at the heart
of the tensor Goldstone  construction of \cite{BG2,RT},
to derive new self-dual systems.

These notes are organized as follows. In sect.~2 we review nonlinear
electrodynamics: we define the notion of self-duality 
and state the self-duality equation which has to be satisfied by the 
action. The derivation can be found in Appendix~A. We also discuss various
properties of self-dual nonlinear electrodynamics, e.g. when 
coupled to a complex scalar field. We then proceed with a description 
of the general structure of self-dual Lagrangians, of which the 
Born-Infeld action is but a particular example, with very special
properties, though. In sect.~3 we present, following Refs. 
\cite{GZ1,Z,Ta,AT},
the generalization to a collection of U(1) vector-fields, coupled
to an arbitrary number of scalar fields. Sect.~4, which is based on
Ref. \cite{KT}, is the ${\cal N}=1$ 
supersymmetric version of sect.~2.  
In sect.~5 we discuss properties of the supersymmetric Born-Infeld
action and make contact with the work of Bagger and Galperin \cite{BG}, where 
this action was obtained as a model of partial 
${\cal N}=2\to {\cal N}=1$ supersymmetry breaking.
In the next section we supersymmetrize the analysis of sect.~4. 
In sect.~7 we discuss self-dual models with tensor multiplets.
In sect.~8, we temporarily leave supersymmetry 
and derive the self-duality equations 
and determine the maximal duality group of a $d$-dimensional system 
with $n$ Abelian $(p-1)$-form potentials and $m$ Abelian 
$(d-p-1)$-form potentials, with and without coupling to scalar fields. 
In sect.~9 we turn to ${\cal N}=2$ supersymmetric models. We 
find the duality equation and demonstrate that the ${\cal N}=2$
Born-Infeld action proposed in Ref. \cite{Ket2} is indeed self-dual. 
This action  correctly reduces to the ${\cal N}=1$ Born-Infeld action
when the $(0,1/2)$ part of the $\cN=2$ vector multiplet
is switched off. However, there are in 
fact infinitely many manifestly ${\cal N}=2$ generalization 
of the ${\cal N}=1$ Born-Infeld action with this property \cite{KT}. 
Within the context of the D3-brane world-volume action, one has to 
impose additional properties (beyond self-duality), in particular 
the action should be invariant under translations in the transverse 
directions in the embedding space, or, in other words, it should contain 
only derivatives of the scalar fields. We show that even when allowing 
for nonlinear field redefinitions, the action of Ref. \cite{Ket2,KT} does not
satisfy this property. 
It is therefore not the correct model 
for partial ${\cal N}=4\to {\cal N}=2$ supersymmetry breaking,
based on the ${\cal N}=2$ Goldstone-Maxwell multiplet. 
We should mention that we know of no \`a priori reason why such a 
theory should be automatically self-dual. 
However this is the case for partial breaking of 
${\cal N}=2$ supersymmetry to ${\cal N}=1$. 
In any case, the manifestly $\cN=2$ supersymmetric 
world-volume action of a D3 brane in $d=6$ is still unknown
(as well as the manifestly $(1,0)$ supersymmetric BI action in 
$d=6$, from which it might be derived via dimensional reduction).

As already mentioned, Appendix~A contains the derivation of the 
self-duality equation in the simplest context, namely of pure
nonlinear electrodynamics. 

At the end of the introduction we want to mention that all our 
considerations are classical. The systems we study should be 
considered as effective theories. That they are relevant is 
demonstrated by the appearance of the Born-Infeld action as the 
world-volume action of D-branes \cite{FT,L}. 
However the 
study of nonlinear self-dual systems might also be interesting in 
its own right. 

Any nonlinear theory must possess a dimensionful parameter.
Within the context of (open) 
string theory this is the string scale $\alpha'$. We 
will always set this parameter to unity.

\sect{Self-duality in nonlinear electrodynamics}

We begin with a review 
\cite{GZ1,Z,GR1,GR2,GZ2,GZ3}
of self-dual models of a single U(1) gauge field
with field strength $F_{ab} = \pa_a  A_b - \pa_b  A_a$.
The dynamics of such a model is determined by a nonlinear Lagrangian 
$L(F_{ab}) = -\frac{1}{4} F^{ab}F_{ab} + \cO(F^4)$. 
With the definition\footnote{We are working 
in $d=4$ Minkowski space, where $\tilde{\tilde F}=-F$,
and often use the notation $F \cdot G= F^{ab} G_{ab}$ 
implying $F \cdot \tilde{G}= \tilde{F}\cdot G$ and 
$ \tilde{F} \cdot \tilde{G} = -F \cdot G$.}
\be
\tilde{G}_{ab} (F)~\equiv~
\hf \, \ve_{abcd}\, G^{cd}(F) ~=~ 
2 \, \frac{\pa L(F)}{\pa F^{ab}}~,\qquad  
G(F) =  \tilde{F} + \cO(F^3)~,
\label{tilde-g}
\ee
the Bianchi identity and the equation of motion read
\be
\pa^b \tilde{F}_{ab} = 0~, \qquad \quad
\pa^b \tilde{G}_{ab} = 0~.
\label{bi+em}
\ee
Since these differential equations, satisfied by $F$, have the same form, 
one may consider {\it duality transformations}\footnote{In the case of 
Maxwell's electrodynamics, the field strength
transforms into its Hodge dual $\tilde{F}$, hence the 
name `duality transformations'.}   
\bea
 \left( \begin{array}{c} G'(F')  \\  F'  \end{array} \right)
~=~  \left( \begin{array}{cc} ~a~& ~b~ \\ ~c~ & ~d~ \end{array} \right) \;
\left( \begin{array}{c} G(F) \\ F  \end{array} \right) ~, \qquad \quad
\left( \begin{array}{cc} ~a~& ~b~ \\ ~c~ & ~d~ \end{array} \right) \in 
{\rm GL}(2, {\Bbb R})~,
\label{GL-duality}
\eea
such that the transformed quantities $F'$ and $G'$ 
also satisfy the equations (\ref{bi+em}). 
{}For $G'$ one should require 
\be
\tilde{G'}_{ab} (F') ~=~ 2 \, \frac{\pa L'(F')}{\pa F'^{ab}}~,
\ee
and the transformed Lagrangian, $L'(F)$, exists 
(in general, $L'(F) \neq  -\frac{1}{4} F \cdot F + \cO(F^4)$)
and can be 
determined
for any ${\rm GL}(2, {\Bbb R})$-matrix entering the transformation 
(\ref{GL-duality}). 
In particular, for an infinitesimal 
duality transformation\footnote{Throughout this paper,
small Latin letters from the beginning of the alphabet
denote finite duality transformation parameters, 
capital letters are used for infinitesimal transformations.}
\bea
 \d \left( \begin{array}{c}  G  \\   F  \end{array} \right)
~=~  \left( \begin{array}{cc} ~A~& ~B~ \\ ~C~ & ~D~ \end{array} \right) \;
\left( \begin{array}{c} G \\ F  \end{array} \right) ~, 
\qquad \quad
\left( \begin{array}{cc} ~A~& ~B~ \\ ~C~ & ~D~ \end{array} \right) \in 
{\rm gl}(2, {\Bbb R})
\eea
one finds
\be
\D L = L'(F) - L(F) = (A+D)\, L(F) -
{1\over2}D \, \tilde G\cdot F\non 
+ \frac{1}{4} B\, F\cdot \tilde{F} -
\frac{1}{4} C\, G\cdot \tilde{G} ~; \label{26}
\ee
{\it c.f.} also sect.~3, eq. (\ref{327}).

The above considerations become nontrivial if one requires
the model to be {\it self-dual}, {\it i.e.}
\be
L'(F) ~= ~ L(F)~.
\ee
The requirement of self-duality implies:\\
(i) only U(1) duality rotations can be consistently 
defined in the nonlinear case, although
Maxwell's case is somewhat special 
(see sect. 3 for details)
\bea
 \left( \begin{array}{c}  G'(F') \\ F'  \end{array} \right)
~=~  \left( \begin{array}{cr} \cos \l ~& ~ 
-\sin \l \\ \sin \l ~ &  ~\cos \l \end{array} \right) \;
\left( \begin{array}{c}  G(F) \\ F  \end{array} \right) ~;
\label{U(1)-duality}
\eea
(ii) the Lagrangian solves the 
{\it self-duality equation} \cite{GR1,GZ2,GZ3}
\be
G^{ab}\, \tilde{G}_{ab} ~+~F^{ab} \, \tilde{F}_{ab} ~=~ 0. 
\label{GZ}
\ee
A derivation of the self-duality equation is presented in Appendix A.

Due to the definition of $G(F)$, 
the self-duality equation severely constrains the possible functional form 
of $L(F)$. Any solution of the self-duality equation defines
a self-dual model.

Self-dual theories possess several remarkable properties: 

\noindent
I. {\it Duality-invariance of the energy-momentum tensor}\\
Given an invariant parameter $g$ in the self-dual theory, 
the observable $\pa L(F,g) /  \pa g$ is duality invariant \cite{GZ1}.
Indeed, using eq. (\ref{var-4}) 
and the duality invariance of $g$, 
one gets
\be
\d \, \frac{\pa }{\pa g} \, L = \frac{\pa}{\pa g} \, \d\, L 
= \hf \l\, \frac{\pa}{\pa g}\left( \tilde{G}\cdot G \right) 
= \hf\l\, \frac{\pa}{\pa g}\left( \tilde{G}\cdot G + \tilde{F}\cdot F
\right) =0~,
\ee
since $F$ is $g$-independent. 
Any self-dual theory can be minimally coupled to the gravitational field 
$g_{mn}$ such that the duality invariance remains intact, 
and $g_{mn}$ does not change under 
the curved-space duality transformations.
Therefore, the energy-momentum tensor is duality invariant.

\noindent
II. SL$(2,{\Bbb R})$ 
{\it duality invariance in the presence of dilaton and axion}\\
Given a self-dual model $L(F)$,
its compact U(1) duality group can be enlarged \cite{GR2,GZ2,GZ3}
to the non-compact 
SL$(2,{\Bbb R})$, by suitably coupling the electromagnetic field to 
the dilaton $\vf$ and axion $a$, 
\be 
\cS = \cS_1 + {\rm i}\, \cS_2 = a + {\rm i} \,{\rm e}^{-\vf}~.
\label{dil-ax}
\ee
Non-compact duality transformations read 
\bea
 \left( \begin{array}{c} G'  \\  F'  \end{array} \right)
=  \left( \begin{array}{cc} a~& ~b \\ c~ & ~d \end{array} \right) \;
\left( \begin{array}{c} G \\ F  \end{array} \right) ~, \qquad 
\cS' = \frac{a\cS +b}{c\cS+d}~, \qquad
\left( \begin{array}{cc} a~& ~b \\ c~ & ~d \end{array} \right) \in 
{\rm SL}(2, {\Bbb R})~,
\label{SL-duality}
\eea
and the duality invariant Lagrangian is 
\be
L(F, \cS, \pa \cS) ~=~ 
L(\cS, \pa \cS)
\,+\,  L( \sqrt{\cS_2}\, F) \,+\, \frac{1}{4}\,\cS_1 F \cdot \tilde{F}~.    
\label{3-dil-ax}
\ee
with $L(\cS, \pa \cS)$ the  SL$(2, {\Bbb R})$ invariant Lagrangian
for the scalar fields,
\be 
L(\cS, \pa \cS) ~=~  \frac{\pa {\bar \cS} \, \pa \cS }
{(\cS - {\bar \cS})^2} ~. \label{kinetic}
\ee 
A derivation of the self-dual model (\ref{3-dil-ax}) 
will be described in sect. 3.

\noindent
III. {\it Self-duality under Legendre transformation}\\
What is usually meant by `{\it duality transformations}' 
in field theory,  
more precisely for models of gauge differential forms
of which  electrodynamics is one example, are 
Legendre transformations. We now show that any system which solves the 
self-duality equation is automatically invariant under 
Legendre transformation.

Let us recall the definition of Legendre transformation 
in the case of a generic model of nonlinear electrodynamics
specified by $L(F)$.
One associates with $L(F)$  an auxiliary model 
$L(F, F_{\rm D})$ defined by
\be
L(F, F_{\rm D}) = L(F) -\hf \, 
F \cdot \tilde{F}_{\rm D}~,
\qquad \quad F_{\rm D}{}^{ab} = \pa^a A_{\rm D}{}^b 
- \pa^b A_{\rm D}{}^a~.
\ee
$F$ is now an unconstrained antisymmetric tensor 
field, $A_{\rm D}$ a Lagrange multiplier field  
and $F_{\rm D}$ the dual electromagnetic field.
This model is equivalent to the original one. Indeed, 
the equation of motion for $A_{\rm D}$ implies
$\pa_b \tilde{F}{}^{ab}=0$ 
and therefore the second term in $L(F, F_{\rm D})$
is a total derivative,  
that is $L(F, F_{\rm D})$ reduces to $L(F)$.
On the other hand, one can first consider the equation of motion
for $F$:
\be
G(F) ~=~ F_{\rm D}~.
\ee
It is solved by expressing $F$ as a function of the dual 
field strength, $F = F (F_{\rm D})$. 
Inserting this solution into $L(F, F_{\rm D})$, 
one gets the dual model
\be
L_{\rm D} (F_{\rm D}) 
\equiv  \Big( L(F)
-{1\over2}F\cdot \tilde F_{\rm D} \Big)\, \Big|_{ F=F(F_{\rm D})} ~.
\label{legendre}
\ee 
It remains to show that 
for any solution $L$ of the self-duality equation, 
its Legendre transform $L_{\rm D}$ satisfies:
\be
L_{\rm D} (F) ~=~L(F)~.
\label{L=L}
\ee
It follows from the results of Appendix A 
that the combination $L- {1\over 4} \;F \cdot \tilde{G}$
is invariant under arbitrary duality rotations, {\it i.e.}
\be 
L(F) - \frac{1}{4}\, F \cdot \tilde{G}(F) 
= L(F') - \frac{1}{4}\, F' \cdot \tilde{G}'(F') ~.
\ee
{}For a finite U(1) duality rotation (\ref{U(1)-duality})
by $\l = \p /2$ this relation reads
\be
L(F) - \frac{1}{2}\, F \cdot \tilde{F}_{\rm D} 
~=~ L(F_{\rm D})  ~,\qquad \quad 
F_{\rm D}~ \equiv ~ G(F)~. 
\label{L=L-2}
\ee
Comparing with (\ref{legendre}) this proves (\ref{L=L}).

Let us turn to a more detailed discussion of the self-duality
equation (\ref{GZ}). Since in four dimensions 
the electromagnetic field has only two independent 
invariants
\be
\a = \frac{1}{4} \, F^{ab} F_{ab}~, \qquad \quad
\b = \frac{1}{4} \, F^{ab} \tilde{F}_{ab} ~,
\ee
its Lagrangian $L(F_{ab})$ can be considered as a 
real function of one complex variable
\be
L(F_{ab})= L(\o , \bar{\o} )~, \qquad \quad
\o = \a + {\rm i} \, \b~.
\label{omega}
\ee
The theory is parity invariant iff $L(\o , \bar{\o} )
= L( \bar{\o}, \o )$. 

One calculates $\tilde{G}$ (\ref{tilde-g}) to be 
\be
\tilde{G}_{ab} = \Big( F_{ab} +{\rm i}\, \tilde{F}_{ab}\Big)
\, \frac{\pa L}{\pa \o} ~+~
\left( F_{ab} - {\rm i} \, \tilde{F}_{ab} \right)\, 
\frac{\pa L}{\pa \bar{\o}}~,
\ee
and the self-duality equation (\ref{GZ}) takes the form
\be 
{\rm Im}\;  \left\{ \o - 4\, \o\, 
\left( \frac{\pa L}{\pa \o} \right)^2 \right\} = 0~.
\label{GZ2}
\ee
In the literature one finds alternative forms of the self-duality 
equation  \cite{GR1,GZ3} but it is eq. 
(\ref{GZ2}) which turns out to be most 
convenient for supersymmetric generalizations.
If one splits $L$ into the sum of Maxwell's part 
and an interaction, 
\be
L = -\hf \, \Big( \o + \bar{\o} \Big) ~+~
L_{{\rm int}}~, \qquad \quad 
L_{{\rm int}} = \cO ( \o ^2)~,
\label{con1}
\ee
(\ref{GZ2}) becomes a condition on $L_{\rm int}$: 
\be 
{\rm Im}\;  \left\{ \o \, \frac{\pa L_{{\rm int}} }{\pa \o}
-  \o\, 
\left( \frac{\pa L_{{\rm int}}   }{\pa \o} \right)^2 \right\} = 0~.
\label{GZ3}
\ee
We restrict $L_{{\rm int}}$ to a real analytic function of
$\o$ and $\bar \o$. Then, every solution of eq. (\ref{GZ3})
is of the form\footnote{In the Euclidean formulation of
self-dual theories, it is
the form (\ref{con2}) of $L_{{\rm int}} $
which allows for (anti)self-dual 
solutions $\tilde{F} = \pm F$ \cite{GH}.} 
\be
L_{{\rm int}} (\o, \bar{\o} ) ~=~
\o \, \bar{\o} \; \L (\o, \bar{\o} )~,
\qquad \quad \L = {\rm const} ~+~ \cO (\o)~,
\label{con2}
\ee
where $\L$ satisfies  
\be
{\rm Im}\;  \left\{ \frac{\pa (\o \, \L) }{\pa \o}
- \bar{\o}\, 
\left( \frac{\pa (\o \, \L )  }{\pa \o} \right)^2 \right\} = 0~.
\label{GZ4}
\ee
Note that for any solution 
$L_{{\rm int}} (\o, \bar{\o} )$ of 
(\ref{GZ3}), or any solution $\L (\o, \bar{\o} )$  
of (\ref{GZ4}), the functions
\be 
\hat{L}_{{\rm int}} (\o, \bar{\o} ) =
\frac{1}{g^2} \, L_{{\rm int}} (g^2 \,\o, 
g^2 \, \bar{\o} )~, \qquad
\hat{\L} (\o, \bar{\o} ) = g^2 \,
\L (g^2 \,\o, g^2 \, \bar{\o} )
\ee
are also solutions for arbitrary real parameter $g^2$.

In perturbation theory one looks for a parity invariant solution
of the self-duality equation by considering the Ansatz
\be 
\L (\o , \bar \o ) ~=~ \sum_{n=0}^{\infty} ~
\sum_{p+q =n} C_{p,q} \; \o^p {\bar \o}^q~,
\qquad \quad C_{p,q}=C_{q,p} ~\in~ {\Bbb R}~,
\ee 
where $n=p+q$ is
the level of the coefficient $C_{p,q}$.
It turns out that for odd level
the self-duality equation uniquely expresses all 
coefficients recursively.
If, however, the level is even, the self-duality equation 
uniquely fixes the level-$n$ coefficients
$C_{p,q}$ with $p \neq q$ through those at lower levels,
while $C_{r,r}$ remain undetermined.
This means that a general solution of the self-duality equation
involves an arbitrary real analytic function of 
one real argument, $f(\o \bar \o )$. 
  
There are a few exact solutions of the self-duality
equation known, the most prominent one being the BI Lagrangian
\cite{BI}
\bea
L_{\rm BI} &=& \frac{1}{g^2} \Big\{\;
1 - \sqrt{- \det (\eta_{ab} + g F_{ab} )} \;
\Big\} \non \\
&=&
 \frac{1}{g^2} \Big\{ \;
1 - \sqrt{1 + g^2 (\o + \bar \o )  
+{1 \over 4}g^4 (\o - \bar \o )^2 } \;
\Big\}~, \non \\
\L_{\rm BI}  &=&
\frac{g^2 }{ 1 + 
\hf g^2(\o + \bar \o )
+ \sqrt{1 + g^2 (\o + \bar \o )  
+\frac{1}{4}g^4 (\o - \bar \o )^2 }}~,
\label{bi-lag}
\eea
with $g$ the coupling constant. 
In the limit $g \to 0$, $L_{\rm BI}$ reduces 
to the Maxwell Lagrangian.
Some other exact solutions of the self-duality equation
were constructed in Ref. \cite{Kam}.

It is worth noting that the BI Lagrangian 
can be given in the form \cite{BG,RT}
\be
L_{\rm BI} =- \hf \,( \chi + {\bar \chi} )~,
\ee
where the complex field $\chi$ is a functions 
of $\o$ and $\bar \o$ which satisfies  the nonlinear constraint
\be 
\chi +  \hf \, g^2  \chi {\bar \chi } - \o ~=~0~.
\ee
As will be discussed below, this form of the BI Lagrangian 
admits nontrivial generalizations \cite{BMZ,ABMZ}.

We close this section with a comment. While we have limited our discussion
to Lagrangians which depend on $F$ but not on its derivatives, 
the latter case can also be treated easily if one considers the action
rather than the Lagrangian and if one defines 
$$
\tilde G[F] ~=~2 \,{\delta S[F] \over\delta F}~,
$$ 
{\it etc.}. This procedure
is mandatory when we treat supersymmetric models. 

\sect{Theory of duality invariance I: 
non-supersymmetric models}

This section has mainly review character. 
We discuss the theory of duality invariance 
of non-supersymmetric models with Abelian gauge fields \cite{GZ1,Z,Ta,AT},
coupled to scalar and antisymmetric tensor fields.
Supersymmetric models will be treated in 
sects.~4-6.

\subsection{Fundamentals}
We consider a theory of $n$ Abelian gauge fields  
coupled to matter fields $\f^\m$. The gauge fields 
enter the Lagrangian only via their field strengths
$F^i_{ab}$, where $i=1,2,\dots,n$,
\be
L ~=~ L( F^i_{ab},\, \f^\m ,\, \pa_a \f^\m ) ~\equiv~ L(\varphi)~.
\ee
As in sect.~2, we introduce the 
dual fields
\be
\tilde{G}^i_{ab} (\varphi)~\equiv~
2 \, \frac{\pa L(\varphi)}{\pa F^{i \, ab}}
\label{B-G}
\ee
which arise in the equations of motion $\pa^b  \tilde{G}^i_{ab} = 0$
for the gauge fields. 

Our aim is to analyze the general conditions  
for the equations of motion (including the Bianchi identities)
of the theory to be invariant under
infinitesimal duality transformations
\bea
 \d \left( \begin{array}{c} G  \\  F  \end{array} \right)
&=&  \left( \begin{array}{cc} ~A~& ~B~ \\ 
~C~ & ~D~ \end{array} \right)  
\left( \begin{array}{c} G \\ F  \end{array} \right) ~, \qquad \quad
\d \f^\m = \x^\m (\f)~.
\label{B-dualtr} 
\eea
Here $A,~B, ~C$ and $D$ are real constant $ n \times n$ matrices, 
and $\x^\m$ are some unspecified functions of the matter fields.
The variation $\d G$ is understood as follows
\be
\d G = G'(\varphi') - G(\vf)~, \qquad
\tilde{G}' (\vf ')=
2 \, \frac{\pa L(\vf ')}{\pa F'} = 
2\, \frac{\pa L(\vf )}{\pa F'} + 
2\, \frac{\pa }{\pa F} \, \d L  ~,
\label{g-var-B}
\ee
where 
\be
\d L ~=~ L(\vf') - L(\vf)~.
\ee

Using the definitions $F' = F + C \,G + D\, F$ and $\f' = \f + \x(\f)$
of the transformed fields, 
one can express the derivative $\pa / \pa F'$ in (\ref{g-var-B})
in terms of those w.r.t. the original fields. 
This gives  
\be
\d \tilde{G}^i_{ab} = 2 \, \frac{\pa }{\pa F^{i\,ab}} \, \d L 
- C^{jk} \,\tilde{G}^j \cdot \frac{\pa G^k}{\pa F^{i\,ab}}
- D^{ji} \, \tilde{G}^j_{ab}~, 
\ee
where we have used the definition (\ref{B-G}).
The latter variation should coincide with $\d \tilde{G}$
that follows from (\ref{B-dualtr}) and their consistency 
is equivalent to the relation 
\bea
& &\frac{\pa }{\pa F^{i\,ab}} \, \left[ 2 \d L 
-\hf B^{jk} \, F^j \cdot \tilde{F}^k 
-\hf C^{jk}\,  G^j \cdot \tilde{G}^k \right] \non \\
& &=  2\Big (A^{ij} + D^{ji} \Big) \frac{\pa L}{\pa F^{j\,ab}} 
+\hf \Big (B^{ij} - B^{ji} \Big) \tilde{F}^j_{ab}
+\hf \Big (C^{kj} - C^{kj} \Big) 
\tilde{G}^j \cdot \frac{\pa G^k}{\pa F^{i\,ab}}~.
\eea
Here the left-hand side is a partial 
derivative of some function with respect to $F$. 
The right-hand side satisfies the same property iff
\be
D + A^{\rm T}= \k \, {\bf 1}~, \qquad
B^{\rm T} = B~, \qquad C^{\rm T} = C~,
\label{B-mat}
\ee
for some real $\k$.
As a result, we find 
\be
\frac{\pa }{\pa F^i} \, \left[  \d L 
-\frac{1}{4} B^{jk} \, F^j \cdot \tilde{F}^k 
-\frac{1}{4} C^{jk}\,  G^j \cdot \tilde{G}^k 
-\k \, L
\right] =0~.
\label{B-cc1}
\ee
This relation expresses the fact that the Bianchi identities
and equations of motion of the gauge fields 
are invariant under the duality transformation
(\ref{B-dualtr}), (\ref{B-mat}).

Now let us turn to the transformation of the matter equation
of motion:
\be
E_\m ~ =~ \frac{\d }{\d \f^\m} \, S[F, \f]
~=~ \Big( \frac{\pa }{\pa \f^\m} 
- \pa_a \,\frac{\pa }{\pa (\pa_a \f^\m)} \Big) \,L~. 
\ee
By definition, its variation reads (it is simpler to work with 
the action)
\bea
\d E &=& \frac{\d }{\d \f'} \, S[F', \f'] - 
\frac{\d }{\d \f} \, S[F, \f] \non \\
&=&
\frac{\d }{\d \f'} \, S[F, \f] + 
\frac{\d }{\d \f} \, \d S~.
\eea
Using $F' = F + C \,G + D\, F$ and $\f' = \f + \x(\f)$
one can express the derivative $\d / \d \f'$ in the second line
in terms of those w.r.t. the original fields.
This leads to 
\be 
\d E_\m ~=~ \frac{\d }{\d \f^\m} \, \left[
\d S - \frac{1}{4} \int {\rm d}^4x \,
 C^{ij}  \tilde G^i \cdot {G}^j \right]
~-~\frac{\pa \x^\n}{\pa \f^\m} \, E_\n~.
\ee
{}From here it is clear that $E_\m$ will transform 
covariantly under duality transformations, 
\be 
\d E_\m ~=~
-\frac{\pa \x^\n}{\pa \f^\m} \, E_\n~,
\ee
if we require
\be
\frac{\d }{\d \f^\m} \, \left[
\d S - \frac{1}{4} \int {\rm d}^4x \,
 C^{jk}  G^j \cdot \tilde{G}^k \right] ~=~ 0~.
\label{B-cc2}
\ee

The relations (\ref{B-cc1}) and (\ref{B-cc2})
are compatible with each other provided $\k = 0$
and hence
\be 
\d L ~=~\frac{1}{4} B^{ij} \, F^i \cdot \tilde{F}^j 
+\frac{1}{4} C^{ij}\,  G^i \cdot \tilde{G}^j ~.
\label{B-L-var}
\ee
It is easy to check that the combination
(the `interaction Hamiltonian') 
$L- {1 \over 4} F^i \cdot \tilde{G}^i $ is 
duality invariant,
\be
\d \Big( L- {1 \over 4}\, F^i \cdot \tilde{G}^i \Big)
~=~0~.
\ee
Eq. (\ref{B-L-var}) can be rewritten in an equivalent, 
but more useful, form if one directly varies $L$ as a
function of its arguments. This leads to 
the self-duality equation
\be 
\d_\f L ~=~ \frac{1}{4} B^{ij} \, F^i \cdot \tilde{F}^j 
-\frac{1}{4} C^{ij}\,  G^i \cdot \tilde{G}^j
+ \hf A^{ij}\,F^i \cdot \tilde{G}^j~,
\label{3-s-d-e}
\ee
where
\be
\d_{\f} L = \left( \x^\m \, \frac{\pa}{\pa \f^\m}
+ (\pa_a \f^\n) \, \frac{\pa \x^\m}{\pa \f^\n}\,
\frac{\pa}{\pa (\pa_a \f^\m)} \right)L~.
\label{var-mat}
\ee

Since $\k = 0$, the condition (\ref{B-mat}) on the matrix
parameters in (\ref{B-dualtr}) can be rewritten
in matrix notation as
\be 
X^{\rm T}\, \O ~+~ \O \, X ~=~0~,
\label{B-mat-1}
\ee
where
\bea
 X  
&=&  \left( \begin{array}{cc} ~A~& ~B~ \\ 
~C~ & ~D~ \end{array} \right)  ~,
\qquad 
\O ~=~ 
\left( \begin{array}{cr} ~0~& ~-{\bf 1}~ \\ 
~{\bf 1}~ & ~0~ \end{array} \right)  ~.
\label{B-mat-2}
\eea
We conclude that Sp($2n, {\Bbb R}$) is the maximal 
group of duality transformations, although in 
specific models
the duality group $G$ may actually be smaller. 
It should be pointed out that Sp($2n, {\Bbb R}$)
or its non-compact subgroup $G$ may appear as the group of 
duality symmetries if the set of matter fields $\f^\m$
include scalar fields parameterizing the coset space 
$G/H$, with $H$ the maximal compact subgroup of $G$ 
(see \cite{GZ1,AT} for a more detailed discussion).
Any self-dual theory without matter, $L(F)$, can be 
understood as a self-dual 
model with matter, $L(F, \f, \pa \f)$,    
with the matter fields frozen, $\f (x) =\f_0 \in G/H$.
The duality transformations preserving this background 
must thus be a subgroup of U$(n)$, the maximal 
compact subgroup of Sp($2n,{\Bbb R}$). 
If one treats the matter fields $\f^\m$ as coupling 
constants, then non-compact duality transformations
relate models with different coupling constants.
It is worth recalling that for the maximal compact subgroup of 
Sp($2n, {\Bbb R}$) the relations (\ref{B-mat-1})
and (\ref{B-mat-2}) should be supplemented by 
$X^{\rm T} = - X$ and hence
\be 
D=A~, \quad C = - B~, \quad A^{\rm T} = - A~, 
\quad B^{\rm T} = B \quad  \Longrightarrow
\quad (B+ {\rm i} A)^\dagger = (B+ {\rm i} A)~.
\label{B-mat-3}
\ee

\subsection{U(n) duality invariant models}
Let us analyze the conditions of self-duality 
for pure gauge theories with maximal duality group U($n$).
Because of (\ref{B-mat-3}) and since $\d_\f L =0$
in the absence of matter, 
the self-duality equation (\ref{3-s-d-e})
reduces to \cite{AT,ABMZ}
$$
 B^{ij} \,( F^i \cdot \tilde{F}^j 
+  G^i \cdot \tilde{G}^j)
~+~2  A^{ij}\,F^i \cdot \tilde{G}^j~=~0.
$$
Since the matrices $A$ and $B$ satisfy eq. (\ref{B-mat-3})
and otherwise arbitrary, the latter relation leads to   
the self-duality equations
\bea
G^i \cdot \tilde{G}^j + F^i \cdot \tilde{F}^j & =&  0~,  \label{B-de1}
\\
( F^i \cdot  \frac{\pa }{\pa F^j}   - 
F^j \cdot  \frac{\pa }{\pa F^i} ) \,L
 & =& 0~.
\label{B-de2}
\eea
The first equation is a natural generalization of the
self-duality equation (\ref{GZ}).
The second equation requires manifest SO($n$) invariance 
of the Lagrangian when $F^i$ transforms in the 
fundamental representation of SO($n$).

The U$(n)$ duality invariant models possess quite 
remarkable properties. In particular, they are self-dual
under a Legendre transformation which acts on a single 
Abelian gauge field while keeping the other $n-1$ fields invariant.
The proof is similar to that given in sect.~2.
Another property is that any U$(n)$ duality invariant model
can be lifted to a model with the maximal 
non-compact duality symmetry Sp($2n, {\Bbb R}$)
by coupling the gauge fields to scalar fields
$\f^\m$ parameterizing the quotient space
Sp($2n, {\Bbb R})\,/\,{\rm U}(n)$ \cite{GZ2,GZ3,AT}.
The case $n=1$ will be discussed in the next subsection.

Nonlinear U$(n)$ duality invariant models with $n>1$
were first constructed in \cite{BMZ,ABMZ} as a generalization 
of the special algebraic representation for the BI action 
reviewed in sect.~2. The Lagrangian reads
\be
L =- \hf \,{\rm tr}\,( \chi + {\bar \chi} )~,
\ee
where the complex $n \times n$ matrix $\chi$
is  a function 
of $F^i$ which satisfies  the nonlinear constraint
\be 
\chi^{ij} +  \hf \,  \chi^{ik} {\bar \chi }^{jk}  
~=~ \o^{ij}~, \qquad 
\o^{ij}~=~\frac{1}{4}
(F^i \cdot F^j + {\rm i} \,F^i
\cdot \tilde{F}^j)~ .
\ee
We refer the reader to \cite{BMZ,ABMZ} for the 
proof of self-duality. The explicit solution of above constraint 
on $\chi$ was provided in Ref. \cite{ABMZ2}.

One might feel uneasy 
with above derivation of the  
self-duality equations (\ref{B-de1}) 
and (\ref{B-de2}) in pure gauge theory $L(F)$ 
as it was essentially based on the relation 
(\ref{B-L-var}) which is valid in the 
presence of matter. 
Without using the matter 
consistency condition (\ref{B-cc2}) we could not 
have set $\k=0$ and, therefore, the variation of $L$
should be
\be 
\d L =
\frac{1}{4} B^{ij} \, F^i \cdot \tilde{F}^j 
+\frac{1}{4} C^{ij}\,  G^i \cdot \tilde{G}^j 
+\k \, L~.
\label{B-L-var-mod}
\ee
However, practically all conclusions turn out to 
remain unchanged if we make use of additional 
physical requirements (the use of matter fields
in the previous consideration simply allows 
to streamline the derivation). 
Let us consider for simplicility 
the case of a single gauge field, $n=1$. 
Then eq. (\ref{B-L-var-mod})
implies ($\k = A+D$) ({\it c.f.} eq. (\ref{26}) with $\Delta L=0$)
\be
\frac{1}{4} B\, F \cdot \tilde{F} -
\frac{1}{4} C\, G \cdot \tilde{G}
= D  \frac{\pa L}{\pa F}  \cdot F 
- (A+D) \, L \label{327}
~.
\ee
Assuming that $L$ is parity even, the expressions
on both sides have different  
parities and should vanish separately
\bea
&& B\, F \cdot \tilde{F} -
 C\, G \cdot \tilde{G} =0~, \label{x1} \\
&& D  \frac{\pa L}{\pa F}  \cdot F 
=(A+D) \, L ~. \label{x2}
\eea
Let us also assume that $L$ reduces to 
Maxwell's  Lagrangian in the weak field
limit, $L = -\frac{1}{4}F \cdot F +\cO(F^4)$,
hence $G = \tilde{F} +\cO(F^3)$, 
$\tilde{G} = - F +\cO(F^3)$, 
and therefore eq. (\ref{x1}) means
\be 
(B+C)\, F \cdot \tilde{F} +\cO(F^4) ~=~0~.
\ee
To the lowest order, this is satisfied iff $B=-C$.
Eq. (\ref{x2}) means that $L(F)$ is a homogeneous function
provided $D \neq 0$. This equation requires $D=A$
if $L = -\frac{1}{4}F \cdot F $ and $D=A=0$ 
otherwise. We see that only U(1) duality rotations are possible 
in nonlinear electrodynamics, while in Maxwell's theory
one can also allow scale transformations.
The latter are however forbidden if one requires invariance
of the energy-momentum tensor under duality transformations.

\subsection{Coupling to dilaton and axion}
We are going to prove that any U(1) duality invariant 
model $L(F)$ can be uniquely coupled to the dilaton and axion
such that the resulting model $L(F,\cS )$ is 
invariant under
SL$(2,{\Bbb R}) \cong  {\rm Sp}(2, {\Bbb R}) $ 
duality transformations \cite{GR2,GZ2,GZ3}.
This property was stated in sect. 2.

{}Following the notation of subsect. 3.1, the case under 
consideration corresponds to $n=1$ and 
$\f^\m = ( \cS, ~\bar \cS )$. In accordance 
with eq. (\ref{SL-duality}), the infinitesimal 
 transformation of $\cS$ reads
\be
\d \cS ~=~ B + 2A\, \cS - C\, \cS^2~.
\ee
To describe the interaction of the dilaton and axion with 
the gauge field, we assume that 
the total Lagrangian is of the form $L(\cS,\pa\cS)+L(F,\cS)$
where the duality invariant kinetic term was given in (\ref{kinetic}). 
The self-duality equation (\ref{3-s-d-e})
is now equivalent to the following three
equations on $L(F,\cS)$:
\bea
2\cS \, \frac{\pa L}{\pa \cS} + 
2{\bar \cS}\, \frac{\pa L}{\pa \bar \cS} &=&
F \cdot \frac{\pa L}{\pa F} ~, \non \\
\frac{\pa L}{\pa \cS} +\frac{\pa L}{\pa \bar \cS}
&=& \frac{1}{4} F \cdot \tilde{ F} ~, \non \\
\cS^2 \, \frac{\pa L}{\pa \cS} + 
{\bar \cS}^2\, \frac{\pa L}{\pa \bar \cS} &=&
\frac{1}{4} G \cdot \tilde{ G}~.
\eea
Inspection of these equations shows
that $L(F,\cS )$ is
\be
L(F, \cS) ~=~ 
L_( \sqrt{\cS_2}\, F) \,
+\, \frac{1}{4}\,\cS_1 F \cdot \tilde{F}~,    
\label{lag-dil-ax}
\ee
where $L(F)$ solves the self-duality equation (\ref{GZ}).
Since $L(F,\cS )$ is self-dual, the combination
$ L- {1 \over 4}\, F \cdot \tilde{G}$ is duality  
invariant. Its invariance under a finite duality rotation by 
$\p /2$ is equivalent to the fact that the Legendre transform
of the Lagrangian is 
\be
L(F,\cS) - \frac{1}{2}\, F \cdot \tilde{F}_{\rm D} 
~=~ L(F_{\rm D} , -\frac{1}{\cS})  ~,\qquad \quad 
F_{\rm D}~ \equiv ~ G(F)~, 
\ee
{\it c.f.} eq. (\ref{L=L-2}).

\subsection{Coupling to NS \mbox{$B$}-field and RR fields}
Within the context of type IIB string theory, one is interested
in duality-invariant couplings of the  
model (\ref{lag-dil-ax}) 
to the NS and RR two-forms, $B_{ab}$ and $C_{ab}$, 
and the RR four-form, 
$C_{abcd}$ (which are possible bosonic background fields).
{\it E.g.} the self-duality of
the world-volume theory of a D3-brane is inherited from the 
SL($2,\Bbb R$) symmetry of type IIB supergravity
\cite{KO} (see also \cite{D-branes}). 
These fields transform under SL$(2,{\Bbb R})$ as
\bea
\left( \begin{array}{c} C'  \\  B'  \end{array} \right)
&=&  \left( \begin{array}{cc} a~& ~b \\ c~ & ~d \end{array} \right) \;
\left( \begin{array}{c} C \\ B  \end{array} \right) ~, \non \\
\tilde{C_4}' &=& \tilde{C_4} +\frac{1}{4}bd \, B \cdot \tilde{B}
+\hf bc\, B \cdot \tilde{C} 
+\frac{1}{4}ac \,C \cdot \tilde{C}~.
\eea 
The transformation of $\tilde{C_4}$ provides a nonlinear
representation of SL$(2,{\Bbb R})$.\footnote{Note that the combination 
$\tilde C_4-{1\over4}C\cdot\tilde B$ is SL($2,\Bbb R$) invariant.}
In the presence of 
$B_2$, $C_2$ and $C_4$, the Lagrangian (\ref{lag-dil-ax})
is extended to 
\be
L(F, \cS, B, C, \tilde{C_4}) ~=~ 
 L( \sqrt{\cS_2}\, \cF) \,+\, \frac{1}{4}\,\cS_1 \cF \cdot \tilde{\cF}
+ \tilde{C_4} -\hf C\cdot \tilde{\cF}~,    
\label{lag-NS-RR}
\ee
where
\be
\cF_{ab} = F_{ab} + B_{ab} ~.
\ee
The theory is invariant under standard gauge transformations
of the gauge forms $B_2$, $C_2$ and $C_4$.
Moreover, the theory is indeed SL$(2,{\Bbb R})$ 
duality invariant.
Given the set of matters fields  
$\f^\m = (\cS, \bar \cS, B_{ab}, C_{ab}, \tilde{C_4})$
it is an instructive exercise to check that 
the self-duality equation (\ref{3-s-d-e})
is satisfied. 

\sect{Self-duality in \mbox{$\cN$} = 1 supersymmetric 
nonlinear electrodynamics}

Gaillard and Zumino conclude their paper \cite{GZ3} by posing 
the following problem: ``When the Lagrangian is self-dual, it is 
natural to ask whether its supersymmetric extension possesses
a self-duality property that can be formulated in a supersymmetric way.''
The problem was solved in \cite{BinG}  
for the case when the Lagrangian is quadratic in 
the U(1) field strengths coupled to supersymmetric matter. 
The solution in the nonlinear case was obtained in 
\cite{KT} for a single vector multiplet and will be extended 
in the sect.~6 to any number of vector
multiplets coupled to scalar multiplets.
In the present section we are going to review 
the $\cN = 1$ supersymmetric results of \cite{KT}.

Let $S[W , {\bar W}]$ be the action generating the dynamics 
of a single $\cN=1$ vector multiplet. The (anti) chiral superfield
strengths ${\bar W}_\ad$ and $W_\a$,\footnote{Our $\cN=1$
conventions are those of \cite{WB,BK}. In particular, 
$z = (x^a, \q^\a , {\bar \q}_\ad )$ are the coordinates of $\cN=1$ 
superspace, ${\rm d}^8z = {\rm d}^4 x\, {\rm d}^2 \q \,
{\rm d}^2 {\bar \q} $ is the full superspace measure, 
and ${\rm d}^6z = {\rm d}^4 x\, {\rm d}^2 \q $ 
is the measure in the chiral subspace.}  
\be
W_\a = -\frac{1}{4}\, {\bar D}^2 D_\a \, V~, \qquad \quad 
{\bar W}_\ad = -\frac{1}{4}\,D^2 {\bar D}_\ad \, V ~,
\label{w-bar-w}
\ee
are defined in terms of a real unconstrained prepotential $V$. 
As a consequence, the strengths are constrained superfields,
that is they satisfy the Bianchi identity
\be
D^\a \, W_\a ~=~ {\bar D}_\ad \, {\bar W}^\ad~.
\label{n=1bi}
\ee

Suppose that $S[W , {\bar W}] \equiv S [v]$ can be unambiguously
defined\footnote{This is always possible if $S[W , {\bar W}]$ 
does not involve the combination $D^\a \,W_\a$ as an 
independent variable.}
 as a functional of {\it unconstrained}
(anti) chiral superfields ${\bar W}_\ad$ and $W_\a$.
Then, one can define {\it (anti) chiral} superfields
${\bar M}_\ad$ and $M_\a$  as
\be
{\rm i}\,M_\a \,[v]\equiv 2\, \frac{\d }{\d W^\a}\,S[v]
~, \qquad \quad
- {\rm i}\,{\bar M}^\ad \,[v]\equiv 2\, 
\frac{\d }{\d {\bar W}_\ad}\, S[v] ~,
\label{n=1vd}
\ee
with the functional derivatives defined in the standard way
\bea
\d S &=& 
\int {\rm d}^6 z\, \d W^\a \; \frac{\d S}{\d W^\a} + 
\int {\rm d}^6 {\bar z}\, \d {\bar W}_\ad \; 
\frac{\d S}{\d {\bar W}_\ad} ~, \non \\
\frac{\d}{\d W^\a (z)} \, W^\b (z') &=&
\d_\a{}^\b\, \Big(-\frac{1}{4} {\bar D}^2 \Big) \, 
\d^4 (x-x') \, \d^2(\q - \q') \,
\d^2(\bar \q - {\bar \q}')~.
\eea 
The vector multiplet equation of motion 
following from the action $S[W,\bar{W}]$ reads
\be
D^\a \, M_\a ~=~ {\bar D}_\ad \, {\bar M}^\ad~.
\label{n=1em}
\ee

Since the Bianchi identity (\ref{n=1bi}) and the equation of
motion (\ref{n=1em}) have the same functional form,
one may consider, similar to the non-supersymmetric case, 
U(1) duality rotations
\bea
 \left( \begin{array}{c}  M'_\a \,[v'] \\ W'_\a  \end{array} \right)
~=~  \left( \begin{array}{cr} \cos \l ~& ~ 
-\sin \l \\ \sin \l ~ &  ~\cos \l \end{array} \right) \;
\left( \begin{array}{c}  M_\a \, [v] \\ W_\a  \end{array} \right) ~,
\label{N=1dualrot}
\eea
where $M'$ should be
\be
{\rm i}\,M'_\a \,[v']= 2\, \frac{\d }{\d W'^\a}\,S[v']~.
\ee
In order for such duality transformations to be consistently defined,
the action $S[W, \bar W ]$ must satisfy a generalization of the
self-duality equation (\ref{GZ}). 
Its derivation follows essentially the same steps as
described in Appendix A, but with a proper replacement 
of partial derivatives by functional derivatives.
To preserve the definition (\ref{n=1vd}) of $M_\a$ and its 
conjugate, the action should transform under 
an infinitesimal duality rotation  as 
\be
\d S = S[v'] - S[v] =
\frac{\rm i}{4}\,\l \,  \int {\rm d}^6 z\, 
\left\{M^\a M_\a - W^\a W_\a \right\} ~+~ {\rm c.c.}
\ee
On the other hand, $S$ is a functional of $W_\a$ and 
${\bar W}_\ad$ only, and therefore its variation is 
\be 
\d S = \frac{\rm i}{2}\,\l \, \int {\rm d}^6 z\, 
M^\a M_\a  ~+~{\rm c.c.}
\ee
Since these two variations must coincide,
we arrive at the following reality condition
\be 
{\rm Im} \int {\rm d}^6 z\, 
\Big( W^\a W_\a ~+~ M^\a M_\a \Big) ~=~0~.
\label{n=1dualeq}
\ee

In eq. (\ref{n=1dualeq}), the superfield
$M_\a$ was defined in (\ref{n=1vd}), 
and $W_\a$  should be considered 
as an {\it unconstrained} chiral superfields. 
Eq. (\ref{n=1dualeq}) is the condition 
for the $\cN=1$ supersymmetric theory to be 
self-dual.   
We call it 
the $\cN=1$ self-duality equation.

With proper modifications, 
the properties of self-dual theories, which we described in sect.~2,
also hold for $\cN=1$ self-dual models.
In particular, the derivative of the 
self-dual action with respect to an invariant parameter is always
duality invariant.
This implies duality invariance  of the $\cN=1$ supercurrent, 
i.e. the multiplet of the energy-momentum tensor
(see \cite{BK} for a review). 
Duality invariant couplings to the dilaton-axion multiplet 
will be discussed in sect.~6. 
Here we would like to concentrate on 
self-duality under $\cN=1$ Legendre transformation,
defined as follows.
Given a vector multiplet model  $S[W , {\bar W}]$,     
we introduce the auxiliary action 
\be
S[{W} , {{\bar W}}, W_{\rm D}, \bar W_{\rm D}] =
S[{W},{\bar W} ]
-{{\rm i}\over 2} \int {\rm d}^6 z\,{W}^\a W_{{\rm D}\, \a}
+{{\rm i}\over2}\int {\rm d}^6 \bar z \,
{\bar W}_\ad \bar W_{\rm D}{}^\ad ~,
\label{n=1da}
\ee
where ${W}_\a$ is now an unconstrained chiral spinor 
superfield, and $W_{{\rm D}\, \a}$ the dual field strength
\be
W_{{\rm D}\,\a} = -\frac{1}{4}\, {\bar D}^2 D_\a \, V_{\rm D}~,
\qquad \quad 
{\bar W}_{{\rm D}\, \ad }= -\frac{1}{4}\,D^2 {\bar D}_\ad \, V_{\rm D} ~.
\ee
This model is equivalent to the original model, since
the equation of motion for $W_{\rm D}$ implies that
${W}$ satisfies the Bianchi identity (\ref{n=1bi}),
and the action  (\ref{n=1da}) reduces to $S[W , {\bar W}]$.
On the other hand, the equation of motion for ${W}$ 
is $M [{W}, {\bar W}] ~=~  W_{\rm D}$, with $M$ defined 
in (\ref{n=1vd}). Solving this equation,  
${W}= {W}[W_{\rm D}, {\bar W}_{\rm D}] $, 
and inserting the solution back
into the action (\ref{n=1da}), one gets the dual 
model $S_{\rm D}[W_{\rm D} , {\bar W}_{\rm D}]$ or,
what is the same, the Legendre transform of 
$S[W , {\bar W}]$. For all $\cN=1$ self-dual 
theories, $S_{\rm D} =S$. This follows from the fact 
that the combination
\be
S -{{\rm i}\over 4}\int {\rm d}^6 z\, W^\a M_\a
+{{\rm i}\over4}\int {\rm d}^6 \bar z \,{\bar W}_\ad 
{\bar M}^\ad 
\ee
is invariant under arbitrary U(1) duality rotations.

We now  present a family of $\cN=1$ 
supersymmetric self-dual models with actions 
of the general form 
\be 
S ~=~ \frac{1}{4}\int {\rm d}^6z \, W^2 +
\frac{1}{4}\int {\rm d}^6{\bar z} \,{\bar  W}^2 
+  \frac{1}{4}\, \int {\rm d}^8z \, W^2\,{\bar W}^2  \,
\L \Big( \frac{1}{8} D^2\,W^2 \, ,\, \frac{1}{8}
{\bar D}^2\, {\bar W}^2 \Big)~,
\label{gendualaction}
\ee
where $\L(u,\bar u )$ is a real analytic function 
of the complex  variable 
\be 
u ~ \equiv ~ \frac{1}{8} D^2 \, W^2~.
\ee 
Functionals of this type naturally appear as
low-energy effective actions in quantum supersymmetric gauge theories;
by `low-energy action'
we mean here the part of the full effective action
independent of the derivatives of the U(1) field strength $F$. 
In fact, the low-energy effective actions usually have 
the more general form
(see, for instance, \cite{MG,PB,BKT}): 
\be 
S_{\rm eff} ~=~ \frac{1}{4}\int {\rm d}^6z \, W^2 +
\frac{1}{4}\int {\rm d}^6{\bar z} \,{\bar  W}^2 
+   \int {\rm d}^8z \, W^2\,{\bar W}^2  \,
\O \Big( D^2\,W^2 ,
{\bar D}^2\, {\bar W}^2 ,  D^\a \, W_\a \Big)~. 
\label{quancor}
\ee
However, the combination 
$ D^\a \, W_\a$
is nothing but the free equation of motion of the $\cN=1$
vector multiplet. Contributions to effective action, which contain 
factors of the classical equations of motion, are ambiguous. They are
often ignored. It is worth pointing out that there is no unique
way to define the action (\ref{quancor}) as a functional of unconstrained
chiral superfield $W_\a$ and its conjugate 
(what is required in the framework of our approach 
to supersymmetric self-dual theories) 
when $\O$ depends on $ D^\a \, W_\a = {\bar D}_\ad \,{\bar W}^\ad$.

Let us analyze the conditions  for
the model (\ref{gendualaction}) to be self-dual.
One finds 
\be
{\rm i}\,M_\a ~=~ W_\a \,\left\{\;
1 - \frac{1}{4} \,{\bar D}^2 \left[ {\bar W}^2 \left( \L + 
\frac{1}{8}\,D^2 \Big( W^2 \, \frac{\pa \L}{\pa u} \Big) \right) \right]
\; \right\}~.
\ee
Then, eq. (\ref{n=1dualeq}) leads to 
\be
{\rm Im}\, \int {\rm d}^8z \, W^2\,{\bar W}^2  \,
\Big( \G ~-~ {\bar u}\, \G^2
\Big)~=~0~,
\label{prelde}
\ee
where 
\be
\G ~ \equiv ~ \L ~+~ \frac{1}{8}\,
(D^2\,W^2 ) \;
\frac{\pa \L}{\pa u} ~=~
\frac{ \pa (u\,\L)}{\pa u}~. 
\ee
In deriving eq. (\ref{prelde}) 
 we have used the following property of the $\cN=1$ 
vector multiplet:
\be 
W_\a \, W_\b \, W_\g ~=~0~.
\ee
Since the functional relation (\ref{prelde}) 
must be satisfied
 for  arbitrary
(anti) chiral superfields ${\bar W}_\ad$ and
$W_\a$, we arrive at the following differential
equation for $\Lambda(u, \bar u )$:
\be
{\rm Im}\;  \left\{ \frac{\pa (u \, \L) }{\pa u}
- \bar{u}\, 
\left( \frac{\pa (u \, \L )  }{\pa u} \right)^2 \right\} = 0~.
\label{dif}
\ee 
This equation is identical to the self-duality equation (\ref{GZ4}).

To obtain the component form of (\ref{gendualaction}),
one applies the reduction rules
\be 
\int {\rm d}^8z \, U = \frac{1}{16} \int {\rm d}^4x \, 
D^2 {\bar D}^2 \,U \, \big|_{\q =0} ~, \qquad 
\int {\rm d}^6z \, U_{\rm c} = - \frac{1}{4}  
\int {\rm d}^4x \, D^2 \,
U_{\rm c}\, \big|_{\q =0} ~.
\ee
We also introduce the component fields of the $\cN=1$ vector multiplet, 
$\big \{\l_\a , \,{\bar \l}_\ad, \,
F_{ab}, \, D \big\}$, in the standard way \cite{WB,BK}:
\bea
\l_\a (x) & = & W_\a |_{\q =0}~, \non \\
F_{\a \b} (x) &=& -\frac{\rm i}{4} 
( D_\a W_\b + D_\b W_\a )|_{\q = 0}~, \non \\
D(x) &=& -\hf D^\a W_\a |_{\q = 0}~,
\eea
with 
\be
F_{\a \ad \, \b \bd } \equiv (\s^a)_{\a \ad}
(\s^a)_{\b \bd} F_{ab} = 2 \ve_{\a \b} \, 
{\bar F}_{\ad \bd} + 2 \ve_{\ad \bd} \, F_{\a \b} ~.
\ee
Here we are interested only in 
the bosonic sector of the model  and 
therefore set $\l_\a = 0$ in what follows.
Under this assumption one can readily compute the 
component Lagrangian
\be
L (F_{ab}, \,D) ~=~ -\hf ( {\bf u} + {\bar {\bf u}} ) ~+~
{\bf u}  {\bar {\bf u}}\, \L({\bf u}, {\bar {\bf u}})~, \qquad
{\bf u} ~ \equiv ~ \frac{1}{8} D^2 W^2 |_{\q = 0} ~=~ 
\o - \hf D^2~,
\ee
with $\o$ defined in eq. (\ref{omega}).
Since only even powers of the auxiliary field $D$ appear 
in $L$, its equation of motion has 
the solution $D=0$. If we take this solution, the duality 
equation (\ref{dif}) implies that the non-supersymmetric 
model $L(F) = L(F, \, D=0)$ is self-dual.

We arrive at the conclusion: every non-supersymmetric self-dual
model of the type considered in
sect.~2 admits an $\cN=1$ supersymmetric  extension 
which is self-dual under manifestly supersymmetric 
duality rotations. The procedure of constructing such 
a supersymmetric extension is constructive: 
given a self-dual Lagrangian $L(F)$, one should first derive 
$\L (\o, \bar \o )$ defined by eqs. (\ref{con1}) and
(\ref{con2}), and then use this function to generate the action
(\ref{gendualaction}).

\sect{Properties of the \mbox{$\cN$} = 1 supersymmetric 
BI action}

We use the results of sect.~3 to obtain the 
unique $\cN=1$ supersymmetric self-dual extension of the 
BI theory (\ref{bi-lag}). With the use of $\L_{\rm BI}$
one immediately gets
\bea
S_{\rm BI} &=& 
 \frac{1}{4}\int {\rm d}^6z \, W^2 +
\frac{1}{4}\int {\rm d}^6{\bar z} \,{\bar  W}^2 
+  { g^2 \over 4} \,  \int {\rm d}^8z \, \frac{W^2\,{\bar W}^2  }
{ 1 + \hf\, A \, + 
\sqrt{1 + A +\frac{1}{4} \,B^2} }~,
\non  \\
 A &=&   { g^2 \over 8} \, 
\Big(D^2\,W^2 + {\bar D}^2\, {\bar W}^2 \Big)~,
\qquad
B = \, { g^2 \over 8} \, \Big(D^2\,W^2 - 
{\bar D}^2\, {\bar W}^2 \Big)~. \label{bi}
\eea
In what follows, for convenience
 we fix the coupling constant to $g^2=4$.

The above action was first introduced in  \cite{DP,CF}
as a super extension of the BI theory.  
However, only much later 
it was realized that
the theory encodes a remarkably reach structure. 
Bagger and Galperin \cite{BG}, and later Ro\v{c}ek and 
Tseytlin \cite{RT} discovered that (\ref{bi}) 
is the action for a Goldstone multiplet
associated with  $\cN=2 \to \cN=1$ 
partial supersymmetry breaking.
Using a reformulation of (\ref{bi}) with auxiliary superfields, 
Brace, Morariu and Zumino \cite{BMZ}
demonstrated that the theory is invariant under 
U(1) duality rotations. The latter property has turned out to be  
a simple consequence of the approach developed in \cite{KT}
and reviewed in the previous section.
Below  we give a concise review of the results 
of \cite{BG} on partial $\cN=2 \to \cN=1$ supersymmetry
breaking. 

Bagger and Galperin noticed that the Cecotti-Ferrara action 
(\ref{bi}) can be represented in the form
\be
S~=~  \frac{1}{4}\int {\rm d}^6z \, X +
\frac{1}{4}\int {\rm d}^6{\bar z} \,{\bar  X}~,
\label{bi-2}
\ee
where the {\it chiral} superfield $X $ is a functional
of $W$ and ${\bar W}$ such that it satisfies
the nonlinear constraint
\be
X ~+~ \frac{1}{4} \,  X\, {\bar D}^2  \, 
{\bar X}  ~=~ W^2~.
\label{n=1constraint}
\ee  
Indeed, using the action rule
\be
\int {\rm d}^8z \, U = - \frac{1}{4} \int {\rm d}^6z \, 
{\bar D}^2 \,U 
\ee
and the constraint (\ref{n=1constraint}), 
one can rewrite (\ref{bi-2}) in the form 
\be
S =
 \frac{1}{4}\int {\rm d}^6z \, W^2 +
\frac{1}{4}\int {\rm d}^6{\bar z} \,{\bar  W}^2 
+ \hf \,  \int {\rm d}^8z \, X \,{\bar X}~.
\label{4.4}
\ee
Using the constraint (\ref{n=1constraint})
once more, we can represent $X \,{\bar X}$ as 
\be 
 X \,{\bar X} = \frac{W^2 \, {\bar W}^2 }
{(1 + \frac{1}{4} \, {\bar D}^2 {\bar X})
(1 + \frac{1}{4} \, D^2 X)}~.
\ee
Since $W^3 = 0$, on the right-hand side we can 
safely take $D^2 X$ in an effective form 
$D^2 X_{\rm eff}$ 
determined by the equation 
\be 
D^2 X_{\rm eff} = \frac{D^2 W^2}
{1 + \frac{1}{4} \, {\bar D}^2 {\bar X}_{\rm eff}}~.
\ee
Using this in (\ref{4.4}) one reproduces (\ref{bi}).

The dynamical system defined by eqs. (\ref{bi-2})
and (\ref{n=1constraint}) is 
manifestly $\cN=1$ supersymmetric.
Remarkably, it turns out to be invariant 
under  a second, nonlinearly realized, supersymmetry
transformation
\bea 
\d X &=& 2\e^\a \, W_\a ~, \label{1st-var} \\
\d W_\a &=& \e_\a 
+ \frac{1}{4}\, {\bar D}^2 {\bar X}\, \e_\a
+{\rm i}\,  \pa_{\a \ad} X  \, {\bar \e}^\ad ~,
\label{2nd-var}
\eea
with $\e_\a$ a constant parameter. 
Such transformations commute with the first, 
linearly realized, supersymmetry, 
and altogether they generate 
the $\cN=2$ algebra without central charge.
There is a simple way to derive 
the supersymmetry transformations (\ref{1st-var}) and 
(\ref{2nd-var}).
One first observes that the variation
(\ref{1st-var}) leaves the action (\ref{bi-2})
invariant, as a consequence of 
the explicit form of the field strength $W_\a$, 
see eq. (\ref{w-bar-w}). Due to 
(\ref{n=1constraint}), the variation $\d X$ 
must be induced by a variation of $W_\a$ of the form   
\be 
\d W_\a = \e_\a 
+ \frac{1}{4}\, {\bar D}^2 {\bar X}\, \e_\a
+ \hat{\d} W_\a~,
\ee
where 
$\hat{\d} W $ should satisfy
\be
W^\a \, \hat{\d } W_\a = \frac{1}{4} \, X\, 
{\bar D}^2 {\bar W}_\ad  \, {\bar \e}^\ad
= -{\rm i}\, X\, \pa_{\a \ad } W^\a \, 
{\bar \e}^\ad ~.
\ee
Since 
\be 
W^\a \, X ~=~0~,
\ee
the latter relation can be rewritten as follows
\be 
W^\a \, \hat{\d } W_\a = {\rm i}\,W^\a  \, \pa_{\a \ad } X \, 
{\bar \e}^\ad ~,
\ee
 and we thus arrive at the variation (\ref{2nd-var}).
But this is not yet the end of the story, since one  
still  has to check that the variation (\ref{2nd-var})
is consistent with the Bianchi identity (\ref{n=1bi}).
Indeed it is.
However, in sect.~9 we will see that the above
procedure cannot be directly generalized to the case of $\cN=2$ 
supersymmetry.

In \cite{BG} Bagger and Galperin proved that the action (\ref{bi}) 
is self-dual under the $\cN=1$ Legendre transformation.
Their proof is ingenious but rather involved. 
The results of sect.~4 make this property 
obvious. The $\cN=1$ super BI theory (\ref{bi})
is invariant under U(1) duality rotations, 
and therefore it is automatically 
self-dual under the $\cN=1$ Legendre transformation.

\sect{Theory of self-duality II: 
\mbox{$\cN$} = 1 supersymmetric \\models}

In this section we develop a general formalism of duality 
invariance for $\cN=1$ supersymmetric theories of $n$ Abelian 
vector multiplets, described by chiral spinor strengths $W^i_\a$ 
and their conjugates ${\bar W}^i_\ad$, 
in the presence of supersymmetric matter
-- chiral superfields
$\F^\m$ and their conjugates ${\bar \F}^\m$. 
We will use the condensed notation 
$S[v] = S[W^{\a \,i}, {\bar W}^i_\ad, \F^\m ,{\bar \F}^\m]$
for the action functional and, as in sect.~4, introduce
{\it (anti) chiral} superfields
${\bar M}^{\ad \, i}$ and $M^i_\a$  dual to 
${\bar W}^i_\ad$ and $W^{\a \,i}$:
\be
{\rm i}\,M^i_\a \,[v]\equiv 2\, \frac{\d }{\d W^{\a \,i}}\,S[v]
~, \qquad \quad
- {\rm i}\,{\bar M}^{\ad \,i} \,[v]\equiv 2\, 
\frac{\d }{\d {\bar W}^i_\ad}\, S[v] ~.
\ee
To simplify notation, we introduce
\be
M^i \cdot M^j = \int {\rm d}^6 z\, M^{\a \,i} \, M^j_\a~, \qquad
{\bar M}^i \cdot {\bar M}^j =\int {\rm d}^6 {\bar z}\,
{\bar M}^i_\ad \, {\bar M}^{\ad \,j }
\ee
and similarly for superspace contractions of (anti) chiral scalar
superfields.

\subsection{General analysis}
We are interested in determining the conditions for the theory 
to be self-dual under {\it chiral} superfield duality transformations 
\bea
 \d \left( \begin{array}{c} M  \\  W  \end{array} \right)
&=&  \left( \begin{array}{cc} ~A~& ~B~ \\ 
~C~ & ~D~ \end{array} \right)  
\left( \begin{array}{c} M \\ W  \end{array} \right) ~, \qquad \quad
\d \F^\m = \x^\m (\F^\n)~,
\label{5-dualtr} 
\eea
with $\x^\m$ a {\it holomorphic} functions of the chiral matter fields.
Here $A, B, C$ and $D$ are constant real $ n \times n$ matrices; 
these matrices have to be real, since the Bianchi identities
$D^\a W^i_\a = {\bar D}_\ad {\bar W}^{\ad \,i}$
and the equations of motion 
$D^\a M^i_\a = {\bar D}_\ad {\bar M}^{\ad \,i}$ 
are special  reality conditions.

By self-duality we understand the following:\\
I. We require
\be
{\rm i} \,M'[v'] = 2 \frac{\d}{\d W'} \,S[v'] =
2 \frac{\d}{\d W'} \, S[v] + 2 \frac{\d}{\d W} \,\d S~, 
\ee
where 
$\d S = S[v'] - S[v]$.\\
II.  The $\F$-equation of motion 
\be 
E_\m [v] = \frac{\d}{\d \F^\m} S[v]
\ee 
transforms covariantly under duality transformations
\be
\d E_\m = - \frac{\pa \x^\n (\F)}{\pa \F^\m}\, E_\n ~,
\ee
where 
\be
\d E = E'[v'] -  E[v] ~,\qquad
E'[v'] =  \frac{\d}{\d \F'} \,S[v'] =
\frac{\d}{\d \F'} \, S[v] +  \frac{\d}{\d \F} \,\d S~.
\ee

Analysis of the self-duality conditions is similar to 
the non-supersymmetric case described in sect.~3.
The transformation law (\ref{5-dualtr})
and condition I 
are consistent provided
\bea
&& \frac{\d }{\d W^{\a \,i}} \left[
 \d S 
- \frac{\rm i}{4}\, B^{jk} \Big(
W^j \cdot W^k -
{\bar W}^j \cdot {\bar W}^k \Big)
- \frac{\rm i}{4}\, C^{jk} \Big(
M^j \cdot M^k -
{\bar M}^j \cdot {\bar M}^k \Big) 
\right] \non \\
&=&~+  \frac{\rm i}{4}\,( C^{jk}-C^{kj} )
\left(  (  \frac{\d }{\d W^{\a \,i}} M^k )
\cdot M^j - (  \frac{\d }{\d W^{\a \,i}} 
{\bar M}^k )  \cdot {\bar M}^j 
\right) \non \\
&&~+~ \frac{\rm i}{4}\,( B^{ij}-B^{ji} ) W^j_\a
~+~ (D^{ji}+A^{ij}) \, \frac{\d }{\d W^{\a \,j}} S[v]~.
\eea
Since the left-hand side is a total variational derivative, 
the matrices $A,B,C$ and $D$ should be constrained as in eq. 
(\ref{B-mat}). Then, the above relation turns into
\bea
&&\frac{\d }{\d W^{\a \,i}} \Big[
 \d S 
- \frac{\rm i}{4}\, B^{jk} \Big(
W^j \cdot W^k -
{\bar W}^j \cdot {\bar W}^k \Big) \non \\
&& ~~~~~~- \frac{\rm i}{4}\, C^{jk} \Big(
M^j \cdot M^k -
{\bar M}^j \cdot {\bar M}^k \Big)
-\k S[v]
\Big] 
=0~.
\label{5-cc}
\eea
Furthermore, the $\F$-equation of motion 
can be shown to change under duality transformations as
\bea
&& \qquad \qquad 
\d E_\m = - \frac{\pa \x^\n }{\pa \F^\m}\, E_\n \\ 
& & +
\frac{\d }{\d \F^\m} \Big[
 \d S 
- \frac{\rm i}{4}\, B^{jk} \Big(
W^j \cdot W^k -
{\bar W}^j \cdot {\bar W}^k \Big) - \frac{\rm i}{4}\, C^{jk} \Big(
M^j \cdot M^k -
{\bar M}^j \cdot {\bar M}^k \Big)
\Big]\,. \non
\eea
Consequently, condition II is satisfied 
if we  impose the condition
\be 
\frac{\d }{\d \F^\m} \Big[
 \d S 
- \frac{\rm i}{4}\, B^{jk} \Big(
W^j \cdot W^k -
{\bar W}^j \cdot {\bar W}^k \Big) - \frac{\rm i}{4}\, C^{jk} \Big(
M^j \cdot M^k -
{\bar M}^j \cdot {\bar M}^k \Big)
\Big] =0~.
\ee
The latter is consistent with (\ref{5-cc}) provided
$\k=0$.
Therefore,  Sp($2n, {\Bbb R}$) is the maximal duality group
(see sect.~3), and  the action transforms as
\bea
\d S &=& \frac{\rm i}{4} \, \d \, \Big( 
W^i \cdot M^i - {\bar M}^i \cdot {\bar W}^i \Big) \non \\
&=& \frac{\rm i}{4}\, B^{ij} \Big(
W^i \cdot W^j -
{\bar W}^i \cdot {\bar W}^j \Big) + \frac{\rm i}{4}\, C^{ij} \Big(
M^i \cdot M^j -
{\bar M}^i \cdot {\bar M}^j \Big)~.
\label{5-atr}
\eea

Equation (\ref{5-atr}) contains nontrivial information.
The point is that the action can be varied directly, 
\bea
\d S &=& S[v'] - S[v] \non \\ 
     &=& \frac{\rm i}{2} \Big( \d W^i \cdot M^i
- \d{\bar W}^i \cdot {\bar W}^i \Big)
+ \d \F^\m \cdot \frac{\d S}{\d \F^\m} +
\d {\bar \F}^\m \cdot \frac{\d S}{\d {\bar\F}^\m}~,
\eea
and the two results should coincide.
This gives
\bea 
 \d \F^\m \cdot \frac{\d S}{\d \F^\m} &+&
\d {\bar \F}^\m \cdot \frac{\d S}{\d {\bar\F}^\m} \non \\
&=& \frac{\rm i}{4} \, B^{ij} \Big(
W^i \cdot W^j -
{\bar W}^i \cdot {\bar W}^j \Big) - \frac{\rm i}{4}\, C^{ij} \Big(
M^i \cdot M^j -
{\bar M}^i \cdot {\bar M}^j \Big) \non \\
&+& \frac{\rm i}{2} \,A^{ij} \Big(
W^i \cdot M^j -
{\bar W}^i \cdot {\bar M}^j \Big)~.
\label{s-d-e-m}
\eea
This is the self-duality equation in the presence of matter.

In the absence of matter, the maximal duality group is U($n$)
and the transformation parameters in (\ref{s-d-e-m})
are constrained by $B= - C = B^{\rm T}$, $A^{\rm T}= - A$.
If the duality group is U($n$),
then eq. (\ref{s-d-e-m}) leads to the following 
self-duality equations
\bea 
 {\rm Im} \, \Big( W^i \cdot W^j + M^i \cdot M^j \Big) &=& 0~, 
\label{5-sde-1}\\
{\rm Im} \, \Big( W^i \cdot M^j - W^j \cdot M^i \Big) &=&0~. 
\label{5-sde-2}
\eea
Eq. (\ref{5-sde-2}) requires the theory to be invariant 
under SO$(n)$ which acts linearly on $W^i$.
For $n=1$, eq. (\ref{5-sde-1}) reduces to (\ref{n=1dualeq}).

Similar to the non-supersymmetric case \cite{GR2,GZ2,GZ3,AT},
a U$(n)$ duality invariant theory of $n$ Abelian vector 
multiplets can be lifted to an Sp($2n, {\Bbb R}$) duality 
invariant model by coupling the vector multiplets
to scalar multiplets $\F^\m$ parameterizing the quotient space 
Sp($2n, {\Bbb R}) \,/\, {\rm U} (n)$. Below we give a proof for 
$n=1$. 

\subsection{Coupling to the dilaton-axion multiplet}
Our aim here is to couple the system (\ref{gendualaction}),
(\ref{dif}) to the dilaton-axion multiplet $\F$ such that 
the resulting model be ${\rm SL}(2, {\Bbb R})$
duality invariant.
The 
${\rm SL}(2, {\Bbb R})$-transformation of $\F$ 
coincides with the $\cS$-transformation 
(\ref{SL-duality}). Its infinitesimal form is
\be
\d \F = B +2A \,\F -C \, \F^2~.
\ee
The self-duality equation (\ref{s-d-e-m})
is now equivalent to the following requirements
on the action functional 
$ S= S[ W, \F]$:
\bea
2 \F \cdot \frac{\d S}{\d \F} + 
2{\bar \F} \cdot \frac{\d S}{\d \bar \F}  &=&
W \cdot \frac{\d S}{\d W} 
+ \bar W \cdot \frac{\d S}{\d \bar W}  ~, \\
{\delta S\over\delta\F}\cdot 1+{\delta S\over\delta\bar{\F}}\cdot 1
&=& \frac{\rm i}{4}\, \left( W \cdot W - 
{\bar W} \cdot {\bar W} \right) , \\
\F^2 \cdot \frac{\d S}{\d \F} +
{\bar \F}^2 \cdot \frac{\d S}{\d \bar \F}
 &=&
\frac{\rm i}{4} \left( M \cdot M - 
{\bar M} \cdot {\bar M} \right)~.
\eea
We are interested in a solution of these 
equations which for $\F= - {\rm i}$ reduces to
the self-dual system given by eqs. (\ref{gendualaction})
and (\ref{dif}). A direct analysis of the self-duality
equations gives the solution
\bea 
S[W,\F] &=& 
\frac{\rm i}{4}\int {\rm d}^6z \, \F\,W^2 -
\frac{\rm i}{4}\int {\rm d}^6{\bar z} \,
{\bar \F}\,{\bar  W}^2  \label{super-d-a}\\ 
&-&  \frac{1}{16}\, \int {\rm d}^8z \, 
(\F-{\bar \F})^2\, W^2\,{\bar W}^2  \,
\L \Big( \frac{\rm i}{16} (\F- \bar \F )\,D^2\,W^2 
\, ,\, \frac{\rm i}{16}(\F- \bar \F )\,
{\bar D}^2\, {\bar W}^2 \Big)~. \non
\eea
To this action one can add the dilaton-axion 
kinetic term $\int {\rm d}^8z \, K(\F, \bar \F )$,
with $K(\F, \bar \F )$ the K\"ahler potential 
of the K\"ahler manifold
SL($2, {\Bbb R}) \,/\, {\rm U} (1)$.
It is worth pointing out that the dilaton and axion 
(\ref{dil-ax}) are related to $\F$ by the rule
${\bar \cS} = \F|_{\q =0}$.
For the $\cN=1$ super BI action (\ref{bi}), the coupling
to the dilaton-axion multiplet was described in 
\cite{BMZ,ABMZ}.

\subsection{Coupling to NS and RR supermultiplets}
The model (\ref{super-d-a}) can be generalized
by coupling it to supermultiplets containing
the NS and RR two-forms, $B_2$ and $C_2$, 
and the RR four-form, $C_4$.
The extended action is
\be 
S[W,\F, \b, \g, \O] ~=~ S[\cW, \F] 
~+~\left\{  \int {\rm d}^6z \, 
\Big(\O +\hf\, \g^\a \cW_\a \Big)~+~{\rm c.c.} \right\}~,
\label{super-NS-RR}
\ee
where 
\be 
\cW_\a ~=~ W_\a + {\rm i}\,\b_\a~.
\ee
is the supersymmetrization of $F+B$. 
Here $\b_\a$, $\g_\a$ and $\O$ are unconstrained chiral superfields
which include, among their components, the fields $B_2$, $C_2$ 
and $C_4$, respectively. The action is invariant under the following
gauge transformations
\bea 
&\d \b_\a = {\rm i}\, \d W_\a = {\rm i}\,{\bar D}^2 D_\a K_1~, \label{db}\\
&\d \g_\a = {\rm i}\,{\bar D}^2 D_\a K_2,\qquad
\d \O = -\frac{\rm i}{2}\,\cW^\a {\bar D}^2 D_\a K_2~,\\
&\d \O = {\rm i}\, {\bar D}^2 K_3~, \label{6-o}
\eea
with $K_i$ real unconstrained superfields.
Note that $\cW_\a$ is invariant under (\ref{db}).
The transformations of $\b$ and $\g$ imply that 
these superfields describe two tensor multiplets;
{\it c.f.} also sect.~7.
Eq. (\ref{6-o}) implies that all components of $\O$
but $ {\rm Re} \,D^2 \O|_{\q=0}$ can be algebraically 
gauged away; the remaining component transforms as a four-form
and is identified with $\tilde{C_4}$.

The theory (\ref{super-NS-RR}) is ${\rm SL}(2, {\Bbb R})$
duality invariant provided the 
superfields $\b_\a$, $\g_\a$ and $\O$ transform as 
\be
 \left( \begin{array}{c} \g'  \\  \b'  \end{array} \right)
=  \left( \begin{array}{cc} a~& ~b \\ c~ & ~d \end{array} \right) \;
\left( \begin{array}{c} \g \\ \b  \end{array} \right) ~, \qquad
\O' = \O -\frac{\rm i}{4}bd \, \b^2
-\frac{\rm i}{2} bc\, \b \g 
-\frac{\rm i}{4}ac \,\g^2~.
\ee
One can check that the self-duality equation (\ref{s-d-e-m})
is satisfied, with $\F^\m = (\F, \b_\a, \g_\a, \O)$
the set of matter chiral superfields.

\subsection{Example of U(n) self-dual supersymmetric theory}
To conclude, we give an example of U$(n)$ duality 
invariant model \cite{BMZ,ABMZ} describing the dynamics 
of $n$ interacting Abelian vector multiplets $W^i_\a$.
The action is
\be
S~=~  \frac{1}{4}\int {\rm d}^6z \, {\rm tr} \,X +
\frac{1}{4}\int {\rm d}^6{\bar z} \,
{\rm tr} \,  {\bar  X}~,
\ee
where the {\it chiral} matrix superfield $X $ is a functional
of $W^i_\a$ and ${\bar W}^i_\ad$ such that it satisfies
the nonlinear constraint
\be
X^{ij} ~+~ \frac{1}{4} \,  X^{ik}\, {\bar D}^2  \, 
{\bar X}^{jk}  ~=~ W^i \,W^j~.
\label{6-const}
\ee  
The proof of self-duality of this theory 
can be found in \cite{BMZ,ABMZ}.
Obviously, this system is a natural generalization of the 
Bagger--Galperin construction for the $\cN=1$ super BI action, 
which we discussed in sect.~4. 

Since for several vector multiplets $W^3 \neq 0$, 
after solving constraint (\ref{6-const}) 
the action will have a more complicated form than (\ref{bi}).

\sect{Self-dual models with \mbox{$\cN$} = 1 tensor multiplet}

In \cite{BG2,RT} it was shown that partial breaking of $\cN=2$ 
supersymmetry to $\cN=1$ can be described with
the $\cN = 1$ tensor multiplet as the Goldstone multiplet.
The construction of Bagger and Galperin \cite{BG2}
was based on an analogy between the $\cN = 1$ vector 
and tensor multiplets. Here we will pursue the same analogy
to generalize the formalism of sect.~4 to construct nonlinear self-dual
models of the $\cN = 1$ tensor multiplet.

We start with a brief description of the $\cN = 1$ tensor multiplet
\cite{S-G} (see \cite{BK} for more details). The multiplet
is described by a real linear superfield $L$
\be
D^2 L={\bar D}^2 L=0~, \qquad \quad L=\bar L~.
\ee
The general solution of this constraint is 
\be
L = \hf \,(D^\a \eta_\a + {\bar D}_\ad {\bar \eta}^\ad)~, \qquad
\quad {\bar D}_\ad \eta_\a = 0~.
\ee
The chiral spinor superfield $\eta_\a$ is a gauge field defined modulo
transformations
\be
\d \eta_\a  = {\rm i}\,{\bar D}^2 D_\a K~,
\ee
with $K$ a real unconstrained superfield, 
and $L$ is the gauge invariant field strength.
The independent components of $L$ are 
a scalar $\vf = L|_{\q =0}$, 
a Weyl spinor $\j_\a= D_\a L|_{\q =0}$ and its conjugate,
and a vector $\tilde{V}_{\a \ad}= \,\hf \,[ D_\a , {\bar D}_\ad ] L|_{\q =0}$
constrained by $\pa_a \tilde{V}^a = 0$. The constraint means that
$\tilde{V}$ is the dual field strength of an antisymmetric tensor 
field, $\tilde{V}^a = \hf \, \ve^{abcd}\, \pa_b B_{cd}$.

{}For generic models of the tensor multiplet, 
the gauge invariant action is a functional of $L$,
$S[L]$.
Here our consideration will be restricted to 
those models with actions of 
the form $S[\J , \bar \J ]$, where 
\be 
\J_\a = D_\a L~, \qquad \quad D_\b \J_\a = 0~.
\label{t-bi1}
\ee
For example, for the free tensor multiplet we have 
\be 
S_{\rm free} = - \int {\rm d}^8 z\, L^2
= \frac{1}{4}\, \int {\rm d}^6 {\bar z}\,\J^2
+ \frac{1}{4}\, \int {\rm d}^6 z\,{\bar \J}^2~.
\ee
The antichiral spinor $\J_\a$ is a constrained superfield
\be
-\frac{1}{4} \,{\bar D}^2 \J_\a 
+ {\rm i} \,\pa_{\a \ad} {\bar \J}^\ad = 0~.
\label{t-bi2}
\ee
This constraint can be treated as the Bianchi identity.
Its general solution is  (\ref{t-bi1}).
The bosonic components of $\J_\a$ are field strengths
of the zero-form and two-form, $U_a =\pa_a \vf$ and $\tilde{V}_a$,
respectively.

{}For the theory with action $S[\J , \bar \J ]$,
we introduce {\it antichiral} $\U_\a$ and {\it chiral}
${\bar \U}^\ad$ superfields as follows
\be
{\rm i}\,\U_\a \equiv 2\, \frac{\d }{\d \J^\a}\,S
~, \qquad \quad
- {\rm i}\,{\bar \U}^\ad \equiv 2\, 
\frac{\d }{\d {\bar \U}_\ad}\, S ~.
\ee
Then one can check that the equation of motion reads
\be
-\frac{1}{4} \,{\bar D}^2 \U_\a 
+ {\rm i} \,\pa_{\a \ad} {\bar \U}^\ad = 0
\label{t-em}
\ee
which has the same form as the Bianchi identity (\ref{t-bi2}).
Therefore, in  analogy with sect.~4, 
one may consider U(1) duality rotations
\bea
 \left( \begin{array}{c}  \U' \\ \J'  \end{array} \right)
~=~  \left( \begin{array}{cr} \cos \l ~& ~ 
-\sin \l \\ \sin \l ~ &  ~\cos \l \end{array} \right) \;
\left( \begin{array}{c}  \U \\ \J  \end{array} \right) ~.
\eea
The theory proves to be duality invariant iff the
self-duality equation 
\be 
{\rm Im} \int {\rm d}^6 {\bar z}\, 
\Big( \J^\a \J_\a ~+~ \U^\a \U_\a \Big) ~=~0
\ee
is satisfied.

Under duality rotations, the following functional 
\be 
S -{{\rm i}\over 4}\int {\rm d}^6 {\bar z}\, \J^\a \U_\a
+{{\rm i}\over4}\int {\rm d}^6  z \,{\bar \J}_\ad 
{\bar \U}^\ad 
\ee
remains invariant. As in sect.~4, this property implies
self-duality under a superfield Legendre transformation
which is defined by replacing the action $S[\J , \bar \J ]$
with
\be
S[\J , {\bar \J}, \J_{\rm D}, {\bar \J}_{\rm D}] =
S[\J,{\bar \J} ]
-{{\rm i}\over 2} \int {\rm d}^6 {\bar z}\,\J^\a \J_{{\rm D}\, \a}
+{{\rm i}\over2}\int {\rm d}^6 z \,
{\bar \J}_\ad {\bar \J}_{\rm D}{}^\ad ~,
\ee
where $\J_\a$ is now an unconstrained antichiral spinor 
superfield, and $\J_{{\rm D}\, \a}$ the dual field strength
\be
\J_{{\rm D}\, \a} = D_\a L_{\rm D}~, \qquad 
D^2 L_{\rm D}= 0~, \qquad {\bar L}_{\rm D} =L_{\rm D}  ~.
\ee

Since above considerations are very similar 
to those in sect.~4, one can make use of 
the previous results to derive nonlinear
self-dual models of the tensor multiplet.
This is achieved by substituting 
$W^2 \to {\bar \J}^2$ in the action (\ref{gendualaction}).
The results of sec.~6 can also be generalized to the case of 
self-dual systems with several tensor multiplets.

\sect{Self-duality and gauge field democracy}

The general theory of duality invariance in four space-time
dimensions, which was reviewed in 
sect.~3, admits a natural 
higher-dimensional generalization \cite{Ta,AT,ABMZ}. 
In even dimensions $d=2p$, one considers theories
of $n$ gauge $(p-1)$-forms $B^i_{a_1 \dots a_{p-1}}$
coupled to matter 
fields $\f^\m$ such that the Lagrangian is a function
of the field strengths $F^i_{a_1 \dots a_p} =
p \, \pa_{[a_1} B^i_{a_2 \dots a_{p]}}$,
\footnote{Our normalization 
is $\pa_{[a_1}B_{a_2\dots a_p]}
={1\over p!}(\pa_{a_1} B_{a_2\dots a_p}\pm\dots)$} 
matter fields and their derivatives,
$L= L( F,\,\f,\,\pa\f)$.
The action is invariant under the Abelian 
gauge symmetries $B^i\to B^i+d\Lambda^i$ where 
$\Lambda^i$ is any $(p-2)$-form. 
In complete analogy with the four-dimensional case, 
one can introduce duality transformations 
and analyze the conditions for self-duality.
The maximal duality group turns out to depend
on the dimension of the space-time.
For $d=4k$ the maximal duality group is
Sp($2n, {\Bbb R}$), while for $d=4k+2$  
it is O$(n,n)$.
In the absence of matter, the maximal 
duality group is compact: U$(n)$ in 
$d=4k$ dimensions, and 
${\rm O}(n)\times {\rm O}(n)$ for $d=4k+2$.
The fact that the maximal duality group 
depends on the dimension of space-time
was also discussed in \cite{Ta2,CFG,CJLP,J}.

A natural question is what happens to a self-dual
system upon dimensional reduction?
The answer is that one finds a self-dual system
with $(p-1)$-forms and $(d-p-1)$-forms in 
$d$ space-time dimensions, where $d$ is not necessarily even.
We now discuss the general properties of such models.
In $d = 4$ such models also appear as  the bosonic sector
of the self-dual systems of the $\cN=1$ tensor multiplet
we discussed in sect.~7. In fact, the analysis of this 
section was inspired by self-duality of the tensor 
Goldstone multiplet \cite{BG2}.

In $d$ space-time dimensions, we consider a theory
of $n$ gauge $(p-1)$-forms $B^i_{a_1 \dots a_{p-1}}$
and $m$ gauge $(d-p-1)$-forms $C^I_{a_1 \dots a_{d-p-1}}$
coupled to matter fields $\f^\m$. We introduce the gauge invariant 
field strengths
\be
U^i_{a_1 \dots a_p} =
p \, \pa_{[a_1} B^i_{a_2 \dots a_{p]}}~, \qquad \quad
V^I_{a_1 \dots a_{d-p}} =
(d-p) \, \pa_{[a_1} C^I_{a_2 \dots a_{d-p]}}~.
\ee
Without loss of generality, we assume $p < [d/2]$
\footnote{$[.]$ denotes the integer part. The case
$p=[d/2]$ for even $d$ is special and was mentioned at the beginning 
of this section.} 
and then introduce the Hodge-dual of $V$,
\be
\tilde{V}^I_{a_1 \dots a_p}  =
\frac{1}{(d-p)!} \, \ve_{a_1 \dots a_p b_1 \dots b_{d-p}} \,
V^{I\, b_1 \dots b_{d-p} }{}~,
\ee
which is of lower rank than $V$.
The Bianchi identities read
\be
\pa_{[b}\, U^i_{a_1 \dots a_{p}]} = 0~, \qquad \quad
\pa^{b}\, \tilde{V}^I_{b a_1 \dots a_{p-1}} =0~.
\ee

The Lagrangian is required to be a function
of the field strengths, matter fields and their derivatives
\be
L= L( U,\, \tilde{V},\,
\f,\,\pa \f) \equiv L(\varphi)~.
\ee
In terms of the dual variables
\be 
\tilde{G}^i_{a_1 \dots a_p} (\vf)= p!\,
\frac{\pa L (\vf)}{\pa U^{i\, a_1 \dots a_p}}~,
\qquad \quad
H^I_{a_1 \dots a_p} (\vf)= p! \,
\frac{\pa L (\vf)}{\pa \tilde{V}^{I\, a_1 \dots a_p}}~,
\ee
the equations of motion for the gauge fields read
\be
\pa^{b}\, \tilde{G}^i_{b a_1 \dots a_{p-1}} =0~,
\qquad \quad 
\pa_{[b}\, H^I_{a_1 \dots a_{p}]} = 0~.
\ee

The explicit structure of the Bianchi identities 
and equations of motion implies that one may  
consider duality transformations of the form 
\bea
 \d \left( \begin{array}{c} H  \\  U  \end{array} \right)
&=&  \left( \begin{array}{cc} A~& ~B \\ 
C~ & ~D \end{array} \right)  
\left( \begin{array}{c} H \\ U  \end{array} \right) ~, 
\qquad 
\d \left( \begin{array}{c} \tilde{V}  \\  
\tilde{G}  \end{array} \right)
=  \left( \begin{array}{cc} M~& ~N \\ 
R~ & ~S \end{array} \right)  
\left( \begin{array}{c} \tilde{V} \\ \tilde{G}\end{array}\right) ~, \non \\
\d \f^\mu &= &\x^\mu(\f)~.
\label{C-dualtr} 
\eea
Here $A,~B, ~C,~D$ and $M,~N,~R,~S$ are real 
constant matrices, 
and $\x^\m$ are some unspecified functions of the matter fields.
We have suppressed the indices $i,\,I$.
Compatibility of the duality transformations with 
self-duality now imposes the conditions 
\be 
N= C^{\rm T}~, \quad R= B^{\rm T}~, \quad
M + A^{\rm T}=\k {\bf 1}~, \quad
S + D^{\rm T}=\k {\bf 1}~,
\label{C-mrel}
\ee
with $\k$ some real constant, as well as 
the following functional relations
\bea
\frac{\pa }{\pa \tilde{V}^{I\underline{a}}} \, \left[  \d L 
-\frac{1}{p!} B^{Jj} \, \tilde{V}^J \cdot U^j 
-\frac{1}{p!} C^{jJ}\,  \tilde{G}^j \cdot H^J 
-\k \, L
\right] &=&0~, \non \\
\frac{\pa }{\pa U^{i\underline{a}}} \, \left[  \d L 
-\frac{1}{p!} B^{Jj} \, \tilde{V}^J \cdot U^j 
-\frac{1}{p!} C^{jJ}\,  \tilde{G}^j \cdot H^J 
-\k \, L
\right] &=&0~,
\label{C-cc1}
\eea
where we have introduced the notation
\be
\tilde{G}^i \cdot H^J 
=\tilde{G}^{i\, a_1 \dots a_p} 
H^J_{a_1 \dots a_p} 
\equiv 
\tilde{G}^{i\, \underline{a}} H^J_{\underline{a}}~.
\ee
{}Furthermore, the matter equation
of motion transforms covariantly
if one requires
\be
\frac{\d }{\d \f^\m} \, \left[
\d S - \frac{1}{p!} \int {\rm d}^4x \,
 C^{iI}  \tilde{G}^i \cdot H^I 
\right] ~=~ 0~.
\label{C-cc2}
\ee
Eqs. (\ref{C-cc1}) and (\ref{C-cc2}) are then seen to be compatible
if $\k =0$ and if the Lagrangian transforms as
\bea
\d L &=& \frac{1}{p!} B^{Ii} \, \tilde{V}^I \cdot U^i 
+\frac{1}{p!} C^{iI}\,  \tilde{G}^i \cdot H^I \non \\
&=& \d \left( \frac{1}{p!}\, U^i \cdot   \tilde{G}^i \right)
= \d \left( \frac{1}{p!}\, \tilde{V}^I \cdot H^I \right)\,. 
\label{C-de1}
\eea

Since $\k = 0$, eq. (\ref{C-mrel}) means that (\ref{C-dualtr}) becomes 
\bea
 \d \left( \begin{array}{c} H  \\  U  \end{array} \right)
&=&  \left( \begin{array}{cc} A~& ~B \\ 
C~ & ~D \end{array} \right)  
\left( \begin{array}{c} H \\ U  \end{array} \right) ~, 
\qquad 
\d \left( \begin{array}{c} \tilde{V}  \\  
\tilde{G}  \end{array} \right)
=  \left( \begin{array}{rr} -A^{\rm T}~&  C^{\rm T} \\ 
B^{\rm T}  ~ &  -D^{\rm T} \end{array} \right)  
\left( \begin{array}{c} \tilde{V} \\ \tilde{G}  \end{array} \right)\,.
\label{8-rep}
\eea
One easily shows that both variations satisfy the same algebra, namely
gl$(n+m,\Bbb R)$. 
The maximal connected duality group is therefore  
${\rm GL}_0(n+m, {\Bbb R}$). 
The finite form for duality 
transformations is
\bea 
\left( \begin{array}{c} H'  \\  U'  \end{array} \right)
&=& g\,\left( \begin{array}{c} H \\ U  \end{array} \right) ~, 
\qquad 
\left( \begin{array}{c} \tilde{V}'  \\  
\tilde{G}'  \end{array} \right) = 
\left( \begin{array}{cr} {\bf 1} & 0 \\ 
0 & -{\bf 1} \end{array} \right)\, 
(g^{\rm T})^{-1}\,
\left( \begin{array}{cr} {\bf 1}& 0 \\ 
0 & -{\bf 1} \end{array} \right)\, 
\left( \begin{array}{c} \tilde{V} \\ \tilde{G}  \end{array} \right)\,,
\eea
with $g \in {\rm GL}_0(n+m, {\Bbb R}$).
 
Equation (\ref{C-de1}) can be rewritten in a different, more useful, form
if one directly computes $\d L$. This gives the self-duality 
equation
\bea
p!\; \d_{\f} L &=& 
B^{Ii} \, \tilde{V}^I \cdot U^i 
- C^{iI}\,  \tilde{G}^i \cdot H^I \non \\
&+& A^{IJ} \, 
\tilde V^I \cdot H^J
- D^{ij}\, \tilde{G}^i
\cdot U^j ~,
\label{C-de2}
\eea
with $\d_{\f} L$ as in eq. (\ref{var-mat}).

In the absence of matter, the maximal connected duality group 
becomes SO($n+m$), the maximal compact 
subgroup of ${\rm GL}_0(n+m,\Bbb R$); {\it i.e.} 
$A^{\rm T}=-A,\, D^{\rm T}=-D,\,B^{\rm T}=-C$.  
Then, the self-duality  equation (\ref{C-de2})
is equivalent to 
\bea
&& \qquad \qquad \qquad \qquad \qquad
U^i\cdot\tilde V^I+\tilde G^i\cdot H^I=0~, \label{C-de3} \\
&& \left(U^i \cdot \frac{\pa}{\pa U^j}-U^j\cdot\frac{\pa}{\pa U^i}\right)L=0\,,
\qquad 
\left(\tilde V^I\cdot\frac{\pa}{\pa \tilde V^J}
-\tilde V^J\cdot\frac{\pa}{\pa \tilde V^I}\right)L=0\,. \label{C-de4}
\eea
Eq. (\ref{C-de4}) says that the theory is manifestly 
SO$(n)\times {\rm SO}(m)$ invariant.

By analogy with the results of \cite{GZ2,GZ3,AT}, 
any SO($n+m$) duality invariant model $L(U^i, \tilde{V}^I)$
can be lifted to a model with the  
non-compact duality symmetry ${\rm GL}_0(n+m, {\Bbb R}$)
by coupling the gauge fields to scalar fields
$\f^\m$ parameterizing the quotient space
${\rm GL}_0(n+m, {\Bbb R})\,/\,{\rm SO}(n+m)$.

Any SO$(n+m)$ duality invariant model $L(U^i, \tilde{V}^I)$, 
where  $U^i_p = {\rm d} B^i_{p-1}$ and 
$V^I_{d-p} = {\rm d} C^I_{d-p-1}$, with $n \neq 0$ and $m \neq 0$ ,
enjoys self-duality under Legendre transformation
which dualizes two given forms $B^i_{p-1}$ and  $C^I_{d-p-1}$
into a $(d-p-1)$-form and a $(p-1)$-form, respectively.
This is a simple consequence of the duality 
invariance, see sect.~2 for more details.
On the other hand, one can apply a Legendre 
transformation which, say, leaves the gauge $(p-1)$-forms invariant 
but dualizes all gauge $(d-p-1)$-forms into $(p-1)$-forms.
One then obtains a model of $(n+m)$ gauge $(p-1)$-forms.
Remarkably, the SO$(n+m)$ duality symmetry of the original 
model turns into a manifest (linear) SO$(n+m)$ symmetry 
of the dualized model. 
This is a consequence of the self-duality equations
(\ref{C-de3}) and (\ref{C-de4}) and the standard properties
of Legendre transformation.
Therefore, in the models that we have considered 
here, all fields are on the same footing, 
hence the title of this subsection. The SO$(n+m)$ 
duality symmetry is linearly realized if all form 
are of the same degree. 

The self-duality equations (\ref{B-de1}) and (\ref{B-de2})
are difficult to solve. However, 
for (\ref{C-de3}) and (\ref{C-de4}), there exists
a simple scheme to derive their general solution. 
One starts with an SO$(n+m)$ invariant model of $(n+m)$ 
gauge $(p-1)$-forms in $d$ dimensions, 
and then simply dualize $m$
of the fields into gauge $(d-p-1)$-forms  by applying 
the proper Legendre transformation.
The dualized model is invariant under the 
duality transformations.

If $n=m$, there are systems (we will give examples below) which 
are invariant under Sp($2n,\Bbb R$) rather than the maximal 
duality group GL($2n,\Bbb R$). This is the case if the 
matrix parameterizing the infinitesimal transformation 
of $\tilde V$ and $\tilde G$, written in the form
\bea
\d \left( \begin{array}{c} \tilde{G}  \\  
\tilde{V}  \end{array} \right)
&=&  \left( \begin{array}{rr} -D^{\rm T}~&  B^{\rm T} \\ 
C^{\rm T}  ~ &  -A^{\rm T} \end{array} \right)  
\left( \begin{array}{c} \tilde{G} \\ \tilde{V}  \end{array} \right)~,
\eea  
is required to coincide with the transformation of $H$ and $U$
({\it c.f.} (\ref{8-rep})). 
In the absence of matter, the duality group of 
these systems reduces to U$(n)$ 
(see sect. 3) and the self-duality equations
take the form (from now on, we do not distinguish 
between indices $i$ and $I$)
\bea
&& \qquad \qquad 
U^{(i}\cdot\tilde V^{j)}+\tilde G^{(i}\cdot H^{j)}=0~, \label{8-de3} \\
&& \left(U^i \cdot \frac{\pa }{\pa U^j}
+ \tilde V^i\cdot\frac{\pa}{\pa \tilde V^j}
\right)L ~-~ (i\leftrightarrow j)=0~. \label{8-de4}
\eea
Eq. (\ref{8-de4}) means that the Lagrangian is manifestly 
SO$(n)$ invariant. Any U$(n)$ duality invariant model 
can be made  Sp$(2n,{\Bbb R} )$  duality invariant by coupling 
the gauge fields to scalars valued in 
${\rm Sp}(2n,{\Bbb R}) \,/\, {\rm U}(n)$.
For $n=1$ the result reads
\be
L(U, \tilde{V}, \cS) ~=~ 
\frac{1}{p!}\,\cS_1\, U \cdot \tilde{V} ~+~
L(\sqrt{\cS_2} \, U, \sqrt{\cS_2} \, \tilde{V})~,
\ee
with $\cS$ the dilaton-axion field (\ref{dil-ax})
transforming by the rule (\ref{SL-duality})
under the duality group
SL($2, {\Bbb R}$). 

In contrast to U$(1)$ duality invariant models of a single
gauge $(2p-1)$-form in even dimensions $d=4p$, 
U$(1)$ duality invariant models of a gauge $(p-1)$-form
and a gauge $(d-p-1)$-form in arbitrary dimensions $d$
can be considered as reducible, since they involve two 
independent fields. However, the latter models 
possess `self-dual' solutions 
\be
U_{a_1 \dots a_p} = \g\, H_{a_1 \dots a_p}
(U,\tilde{V})~, \qquad 
\tilde{V}_{a_1 \dots a_p}= -\frac{1}{\g}\,\tilde{G}_{a_1 \dots a_p} 
(U,\tilde{V})~,
\ee
with $\g$ a constant parameter. The explicit dependence of $\g$
is dictated by the self-duality equation (\ref{8-de3}).
Such solutions of the equations of motion describe 
the dynamics of a single field.

To conclude, we give an example of a U(1) duality invariant
model. The Lagrangian reads 
\be
L~=~ \frac{1}{p!}~-~  \frac{1}{p!}\,\sqrt{ 
1 + U \cdot U - \tilde{V} \cdot \tilde{V}
-(U \cdot \tilde{V})^2 }~.
\label{C-tm}
\ee
It is easy to check that $L$
solves the self-duality equation (\ref{8-de3}), 
and therefore the theory is U(1) duality invariant.
The theory can be equivalently represented in the form
\be
L =- \frac{1}{2\,p!} \,( \chi + {\bar \chi} )~,
\ee
where the complex field $\chi$ is a functions 
of $U$ and $\tilde{V}$ which satisfies  the nonlinear constraint
\be 
\chi +  \hf \,  \chi {\bar \chi } - \j ~=~0~, \qquad \quad
\j~=~\hf\, (U+{\rm i} \,\tilde{V})^2. \label{824}
\ee
This representation is analogous to that for the BI theory 
described in sect.~2.

The above duality invariant system has a supersymmetric origin.
Let us choose $d=4$ and then $p=1$
is the only interesting choice. The dynamical fields are a
scalar $\vf$ and an antisymmetric gauge field $B_{ab}$
which should enter the Lagrangian only via their field strengths
$U_a = \pa_a \, \vf$ and $\tilde{V}^a = 
\hf \, \ve^{abcd}\, \pa_b \,B_{cd}$. Then, the Lagrangian (\ref{C-tm})
describes the bosonic sector of a model
for partial $\cN=2 \to \cN=1$ supersymmetry breaking 
with the tensor multiplet as the Goldstone multiplet
\cite{BG2,RT}. The antisymmetric gauge field can be dualized into a scalar, 
by applying the appropriate Legendre transformation.
The resulting model is manifestly U(1) invariant 
and it describes  a 3-brane in six dimensions.

Other examples of U(1) duality invariant models of 
the scalar and antisymmetric tensor in four dimensions
can be obtained by considering the bosonic sector
of the self-dual tensor multiplet systems we discussed
in sect.~7. It is worth noting that not all U(1) duality invariant 
models of the scalar and antisymmetric tensor admit a supersymmetric 
extension: the two fields have to appear in the action in the 
combination $\psi$ as defined in (\ref{824}). 
This is in contrast with what we found in self-dual
nonlinear electrodynamics.

Using the results of \cite{BMZ,ABMZ}, the construction just described
can be generalized to derive U$(n)$ duality invariant models
of $n$ gauge $(p-1)$-forms and $n$ gauge $(d-p-1)$-forms
in four dimensions. The Lagrangian is 
\be
L =- \frac{1}{2\,p!} \,{\rm tr}\,( \chi + {\bar \chi} )~,
\ee
where the complex $n \times n$ matrix $\chi$
is  a function 
of $U^i$ and $\tilde{V}^i$ which satisfies  the nonlinear constraint
\be 
\chi^{ij} +  \hf \,  \chi^{ik} {\bar \chi }^{jk}  ~=~
\hf\, (U^i+{\rm i} \,\tilde{V}^i)
\cdot (U^j+{\rm i} \,\tilde{V}^j) ~.
\ee

\sect{\mbox{$\cN$} = 2 duality rotations}
The construction of sect.~4 admits a natural 
generalization to $\cN=2$ supersymmetry \cite{KT}, 
although here much less explicit results have been obtained
so far.
We will discuss the case of one single Abelian gauge multiplet only, 
the generalization to an arbitrary number being straightforward. 
 
We will work in $\cN =2$ global superspace 
${\Bbb  R}^{4|8}$ 
parametrized by 
$\cZ^A = (x^a, \q^\a_i, {\bar \q}^i_\ad) $,
where $i = {\1}, {\2}$. The flat covariant 
derivatives $\cD_A = (\pa_a, \cD^i_\a, 
{\bar \cD}^\ad_i )$ satisfy the algebra
\be
\{ \, \cD^i_\a ,\cD^j_\b \, \} =
\{ \, {\bar \cD}_{\ad i},
{\bar \cD}_{\bd j} \, \} =0~, \qquad
\{ \, \cD^i_\a , {\bar \cD}_{\ad j}\, \} =
-2\,{\rm i}\, \d^i_j \,(\s^a)_{\a \ad}\, \pa_a ~.
\ee
Throughout this section, we will use the
notation:
\bea
\cD^{ij} \equiv \cD^{\a (i} \cD^{j)}_\a 
= \cD^{\a i} \cD^{j}_\a  ~ , 
& \qquad &
{\bar \cD}^{ij} \equiv {\bar \cD}^{(i}_\ad {\bar \cD}^{j)\,\ad}
= {\bar \cD}^{i}_\ad {\bar \cD}^{j\,\ad} \non \\
 \cD^4 \equiv \frac{1}{16}
(\cD^{\underline{1}})^2\,
(\cD^{\underline{2}})^2~, & \qquad &
{\bar \cD}^4 \equiv \frac{1}{16} 
({\bar \cD}_{\underline{1}})^2\,
({\bar \cD}_{\underline{2}})^2  ~.
\eea
An integral over the full superspace 
(with the measure ${\rm d}^{12}\cZ = {\rm d}^4 x\, {\rm d}^4 \q \,
{\rm d}^4 {\bar \q} $) 
can be reduce
to one over the chiral subspace (with the measure
${\rm d}^8\cZ = {\rm d}^4 x\, {\rm d}^4 \q $)
or over the antichiral subspace 
(${\rm d}^8 {\bar \cZ} = {\rm d}^4 x\,
{\rm d}^4 {\bar \q} $):
\be
\int {\rm d}^{12}\cZ \; \cL (\cZ ) ~ =~ 
\int {\rm d}^{8}\cZ \; \cD^4\cL (\cZ )~=~
\int {\rm d}^{8}{\bar \cZ} \; {\bar \cD}^4\cL (\cZ )~.
\ee

The discussion in this section is completely analogous 
to the one presented in the first part of sect.~4. 
We will thus be brief. 
Let $ \cS[\cW , {\bar \cW}]$ be the action describing the dynamics 
of a single $\cN=2$ vector multiplet. The (anti) chiral superfield
strengths ${\bar \cW}$ and $\cW$ 
satisfy the Bianchi identity \cite{gsw}
\be
\cD^{ij} \, \cW ~=~ 
{\bar \cD}^{ij} \, {\bar \cW}~.
\label{n=2bi-i}
\ee
The general solution of the Bianchi identity \cite{Mez},
\be
\cW =  {\bar \cD}^4 \cD^{ij} \, V_{ij}~, \qquad \quad 
{\bar \cW} = \cD^4 {\bar \cD}^{ij} \, V_{ij}
\ee
is in terms of a real unconstrained prepotential $V_{(ij)}$.

Suppose that $\cS[\cW , {\bar \cW}]$ can be unambiguously
defined as a functional of {\it unconstrained}
(anti) chiral superfields ${\bar \cW}$ and $\cW$.
Then, one can define (anti) chiral superfields
${\bar \cM}$ and $\cM$ as
\be
{\rm i}\,\cM \equiv 4\, \frac{\d }{\d \cW}\,
\cS[\cW , {\bar \cW}]
~, \qquad \quad
- {\rm i}\,{\bar \cM} \equiv 4\, 
\frac{\d }{\d {\bar \cW}}\, \cS[\cW , {\bar \cW}]
\label{n=2vd}
\ee
in terms of which the equations of motion are
\be
\cD^{ij} \, \cM ~=~ {\bar \cD}^{ij} \, {\bar \cM}~.
\label{n=2em}
\ee

Again, since the Bianchi identity (\ref{n=2bi-i}) and the equation of
motion (\ref{n=2em}) have the same functional form,
one can consider infinitesimal U(1) duality transformations
\be
\d \cW ~=~ \l \, \cM~, \qquad \quad
\d \cM  ~=~ -\l \, \cW~.
\label{n=2dt}
\ee
The analysis of Appendix A leads to
\bea
\delta \cS &=& -{{\rm i}\over8}\, \l 
\int {\rm d}^8 \cZ \, \Big(\cW^2-\cM^2 \Big)
+{{\rm i} \over 8} \, \l
\int {\rm d}^8\bar{\cZ}\, 
\Big(\bar \cW^2-\bar \cM^2 \Big) \\
&=& {{\rm i}\over4} \, \l \int {\rm d}^8\cZ\,\cM^2
-{{\rm i} \over4} \, \l \int {\rm d}^8\bar\cZ\,\bar \cM^2 \non
\eea
The theory is thus duality invariant
provided the following reality condition is satisfied:
\be 
\int {\rm d}^8 \cZ\, 
\Big( \cW^2 + \cM^2 \Big) ~=~
\int {\rm d}^8 {\bar \cZ}\,
\Big( {\bar \cW}^2  +
{\bar \cM}^2 \Big)\,. 
\label{n=2dualeq}
\ee
Here $\cM$ and ${\bar \cM}$
are defined as in (\ref{n=2vd}), 
and $\cW$ and ${\bar \cW}$ should be considered 
as {\it unconstrained} chiral and antichiral superfields,
respectively. 
Eq. (\ref{n=2dualeq}) serves as our master functional equation
($\cN=2$ self-duality equation)
to determine duality invariant models of the 
$\cN=2$ vector multiplet. 

We remark that, as in the $\cN=0,1$ cases, 
the action itself is not duality invariant, but 
\be
\delta\, \Bigg(\cS-{{\rm i}\over8}\int {\rm d}^8\cZ\, \cM\cW
+{{\rm i}\over 8}\int {\rm d}^8\bar\cZ\,\bar \cM\bar \cW
\Bigg) ~=~0~.
\ee
The invariance of the latter functional under a finite U(1)
duality rotation by $  \pi / 2$, 
is equivalent to the self-duality
of $\cS$ under Legendre transformation, 
\be
\cS[\cW, {\bar \cW} ]
- {{\rm i}\over 4} \int {\rm d}^8 \cZ\, \cW \cW_{\rm D}
+ {{\rm i}\over 4}\int {\rm d}^8 {\bar \cZ} \,
{\bar \cW} {\bar \cW}_{\rm D}
~=~\cS[\cW_{\rm D}, {\bar \cW}_{\rm D} ]~,
\ee 
where $\cW_{{\rm D}}$ is 
the dual chiral field strength,
\be
\cW_{\rm D} =  {\bar \cD}^4 \cD_{ij} \, V_{\rm D}{}^{ij}~, 
\label{dfs}
\ee
with $V_{\rm D}{}^{ij}$ a real unconstrained prepotential.

Apart from the $\cN=2$ Maxwell action
\be
\cS_{\rm free} = 
\frac{1}{8}\int {\rm d}^8 \cZ \, \cW^2 +
\frac{1}{8}\int {\rm d}^8{\bar \cZ} \,{\bar  \cW}^2 ~,
\label{n=2maxwell}
\ee
only one other solution of (\ref{n=2dualeq}) is known \cite{Ket2}:
\be
\cS ~=~
\frac{1}{4}\int {\rm d}^8 \cZ \, \cX +
\frac{1}{4}\int {\rm d}^8{\bar \cZ} \,{\bar \cX}~,
\label{n=2bi}
\ee
where the chiral superfield $\cX$ is a functional 
of $\cW$ and $\bar \cW$ defined via the 
constraint
\be
\cX ~=~ \cX \, {\bar \cD}^4 {\bar \cX} ~+~
\hf \, \cW^2~.
\label{n=2con}
\ee
{}Following \cite{KT}, let us prove that this system provides
a solution of the self-duality equation (\ref{n=2dualeq}).
Under an infinitesimal variation of $\cW$ only, we have 
\bea
\d_{\cW}  \cX &=& \d_{\cW}  \cX \,
{\bar \cD}^4 {\bar \cX} + \cX\,{\bar \cD}^4  
\d_{\cW}  {\bar \cX} + \cW \, \d \cW~, \non \\
\d_{\cW}  {\bar \cX} &=& \d_{\cW}  {\bar \cX} \,
\cD^4 \cX + {\bar \cX}\, \cD^4 \d_{\cW} \cX~.
\eea
{}From these relations one gets
\be
\d_{\cW}  \cX = \frac{1}{1 - \cQ }\,
\left[ \frac{ \cW \, \d \cW}
{ 1 - {\bar \cD}^4 {\bar \cX} } \right] ~, \qquad
\d_{\cW}  {\bar \cX} =
\frac{\bar \cX}{1 - \cD^4 \cX } \, \cD^4 \d_{\cW} \cX~,
\ee
where
\bea
 \cQ = \cP \, {\bar \cP}~, & \qquad &
{\bar \cQ} = {\bar \cP}\, \cP~, \non \\
\cP = \frac{\cX}{1 - {\bar \cD}^4 {\bar \cX} }\,
{\bar \cD}^4~, & \qquad &
{\bar \cP} = \frac{\bar \cX}{1 - \cD^4 \cX } \, \cD^4~.
\eea
With these results, it is easy to compute $\cM$:
\be
{\rm i}\, \cM = \frac{\cW}{1 - {\bar \cD}^4 {\bar \cX} }\,
 \left\{ 1 + {\bar \cD}^4\, {\bar \cP}\, \frac{1}{1 - \cQ }\,
\frac{\cX}{1 - {\bar \cD}^4 {\bar \cX} } +
{\bar \cD}^4\, \frac{1}{1 - {\bar \cQ} }\,
\frac{\bar \cX}{1 - \cD^4 \cX } \right\}~.
\label{M}
\ee 
Now, a short calculation gives
\be
{\rm Im}\,\int {\rm d}^8 \cZ \,\left\{
 \cM^2 
+2\, \frac{1}{1 - \cQ }\,
\frac{\cX}{1 - {\bar \cD}^4 {\bar \cX} } 
\right\} ~=~ 0~.
\label{prom}
\ee
On the other hand, the constraint (\ref{n=2con}) implies
\be
\int {\rm d}^8 \cZ \, \cX - 
\int {\rm d}^8{\bar \cZ} \,{\bar \cX} =
\hf \, \int {\rm d}^8 \cZ \, \cW^2 -   
\hf \,\int {\rm d}^8{\bar \cZ} \,{\bar \cW}^2 ~,
\label{real}
\ee
and hence
\be
\frac{\d }{\d \cW}\,\left\{ \int {\rm d}^8 \cZ \, \cX - 
\int {\rm d}^8{\bar \cZ} \,{\bar \cX} \right\} = \cW~.
\ee
The latter relation can be shown to be equivalent to 
\be
 \frac{1}{1 - \cQ }\,
\frac{\cX}{1 - {\bar \cD}^4 {\bar \cX} }
= \cP \,\frac{1}{1 - {\bar \cQ} }\, 
\frac{\bar \cX}{1 - \cD^4 \cX } + \cX~.
\label{importantrel}
\ee 
Using this result in eq. (\ref{prom}), 
we arrive at the relation
\be
\int {\rm d}^8 \cZ \, \cM^2 - 
\int {\rm d}^8{\bar \cZ} \,{\bar \cM}^2 =
-2 \int {\rm d}^8 \cZ \, \cX
+2 \int {\rm d}^8{\bar \cZ} \,{\bar \cX}
\ee 
which is equivalent, due to (\ref{real}),
to  (\ref{n=2dualeq}).

The dynamical system (\ref{n=2bi}), (\ref{n=2con})
was introduced
in \cite{Ket2} as the $\cN=2$ supersymmetric BI 
action ({\it c.f.} with the similar construction 
for the $\cN=1$ super BI action we described in sect.~3).
Such an interpretation is supported in part by the fact that the theory
correctly reduces to the $\cN=1$ BI in a special $\cN=1$ limit;
we now briefly discuss this issue.

Let us introduce the $\cN=1$ components of the $\cN=2$ 
vector multiplet.
Given an $\cN=2$ superfield $U$, its $\cN=1$ projection 
is defined to be $ U| = U(\cZ)|_{ \q_{ \underline{2} } = 
{\bar \q}^{{\2}} = 0}$. The $\cN=2$ vector multiplet
contains two independent chiral $\cN=1$ components
\be 
\cW | = \sqrt{2}\, \F~, \qquad \quad \cD_\a^{\underline{2}}\, 
\cW |= 2{\rm i}\, W_\a~,\qquad\quad(\cD^{\underline{2}})^2\cW|
=\sqrt{2} \, \bar{D}^2\bar\F~.
\ee
Using in addition that
\be
\int{\rm d}^8\cZ=-{1\over4}\int{\rm d}^6 z \, 
(\cD^{\underline{2}})^2~,
\qquad
\int{\rm d}^{12}\cZ={1\over16}\int{\rm d}^8 z \,
(\cD^{\underline{2}})^2 \, (\bar\cD_{\underline{2}})^2 ~,
\ee
the above definitions imply that
the free $\cN=2$ vector multiplet action (\ref{n=2maxwell})
straightforwardly reduces to $\cN=1$ superfields 
\be
\cS_{\rm free} =   
\int {\rm d}^8z \, {\bar \F} \F +
 \frac{1}{4}\int {\rm d}^6z \, W^2 +
\frac{1}{4}\int {\rm d}^6{\bar z} \,{\bar  W}^2~. 
\ee
If one switches off $\F$,
\be
\F = 0~ \qquad \Longrightarrow \qquad 
( \cD^{\underline{2}} )^2 \cW| = 0~,
\label{f=0}
\ee
one readily observes that
the theory (\ref{n=2bi}), (\ref{n=2con}) 
reduces to the $\cN=1$
BI theory (\ref{bi-2}), (\ref{n=1constraint}).
However, it was shown in \cite{KT}
that there exist infinitely many manifestly $\cN=2$ 
supersymmetric models possessing this very property.
Of course, the specific feature of
the system (\ref{n=2bi}), (\ref{n=2con})
is its invariance under U(1) duality rotations, 
and the requirement of self-duality
severely restricts the class of possible models.
But it turns out that even 
the latter requirement is not sufficient 
to uniquely fix the $\cN=2$ supersymmetric BI action.

The $\cN=2$ supersymmetric BI action is expected to 
describe a single D3-brane in six dimensions
\be
L_{{\rm D3-brane}} ~=~1~ -~\sqrt{-\det \big( \eta_{ab} + F_{ab} 
+ \pa_a {\bar \vf} \pa_b \vf \big)}~.
\ee
Here the complex transverse coordinates $\vf$ of the brane
should, in general, be related to the scalars $\f = \cW|_{\q= 0}$
and the other components 
of the $\cN=2$ vector multiplet by a nonlinear
field redefinition (see, e.g. \cite{T}). Since $L_{{\rm D3-brane}}$ 
is manifestly invariant under constant shifts 
of the transverse coordinates
\be 
\vf (x) ~ \longrightarrow ~ \vf (x) + \s~, 
\ee
the full supersymmetric theory must
also be invariant under such transformations
acting on $\cW$ in a nonlinear way
\be
\cW  (\cZ)~ \longrightarrow ~ \cW (\cZ)+ \s +\cO(\cW, \bar \cW )~.
\label{shifts}
\ee
Moreover, the $\cN=2$ supersymmetric BI action is expected to 
provide a model for partial $\cN=4 \to \cN=2$ supersymmetry 
breaking \cite{BIK}. It means that the action should 
be invariant under nonlinear transformations
\be
\cW  (\cZ)~ \longrightarrow ~ \cW (\cZ)+ \e (\q) +\cO(\cW, \bar \cW )~,
\qquad \e (\q ) = \s +  \e^{\a}_i \, \q^{i}_\a~, 
\label{nonlinearsusy}
\ee
with $\e^{\a}_i$ a constant spinor parameter.
We now demonstrate that the system
(\ref{n=2bi}), (\ref{n=2con}) is not compatible 
even with the simpler transformations (\ref{shifts}).

To start with, it is worth pointing out the following.
When looking for nonlinear symmetry transformations  (\ref{shifts})
or (\ref{nonlinearsusy}), one might first try to 
duplicate the trick\footnote{This course was taken up in \cite{Ket3}.}
 which successfully worked
in the case of the $\cN=1$ supersymmetric BI action
(see sect.~5). Namely, one can introduce the transformation of
$\cX$
\be
\d \cX = \e (\q)\, \cW~,\qquad \quad {\bar \cD}^i_\ad \, \e (\q)=
\cD^{ij}\,\e (\q)=0~, 
\ee
which obviously leaves the action (\ref{n=2bi}) invariant.
But this variation of $\cX $ must be induced by a variation 
of $\cW$ consistent with the constraint (\ref{n=2con}).
A direct analysis shows that the variation $\d \cW$, 
that is derived in this way,
does not satisfy the Bianchi identity (\ref{n=2bi-i}).
The difference from the $\cN=1$ case is simple but crucial:
the $\cN=2$ vector multiplet does not possess any analogue
of the property $W^3=0$, typical for the $\cN=1$ vector multiplet.

We will use the following general Ansatz
\be
\d \cW = \s + \s \,{\bar \cD}^4 \bar \cY + 
{\bar \s}\, \Box \cY~, \qquad  \quad {\bar \cD}^i_\ad \cY = 0
\label{shifts2}
\ee
for symmetry transformations (\ref{shifts}).
The variation is consistent with the Bianchi 
identity (\ref{n=2bi-i}). The chiral superfield
$\cY$ is some unknown functional of $\cW$ and $\bar \cW$.
The precise form of $\cY$ as well as of the 
$\cN=2$ supersymmetric BI action,
$\cS_{\rm BI}$,
should be determined, order by order in perturbation theory,
from three requirements: (i) the action is to be invariant
under transformations (\ref{shifts2});
(ii) the action should solve the self-duality 
equation (\ref{n=2dualeq}); (iii) to order $\cW^4$, the action 
should have the form:
\be
\cS_{\rm BI} = \cS_{\rm free} 
~+~ \cS_{\rm int} ~, \qquad
\cS_{\rm int} = 
{ 1 \over 8} \,  \int {\rm d}^{12} \cZ \, 
\cW^2\,{\bar \cW}^2 ~+~\cO(\cW^6)~.
\ee
This reproduces the known $F^4$ terms in the BI action.\footnote{This 
is the only known superinvariant with this property.}
Direct calculation gives for $\cY$
\bea 
\cY &=& -\hf\, \cW^2  \Big\{ 1 + \frac{1}{2}\, 
{\bar \cD}^4 \,{\bar \cW}^2 
+ \frac{1}{8}\, {\bar \cD}^4 \,({\bar \cW}^2 \, \cD^4 \cW^2)
+ \frac{1}{8}\, ({\bar \cD}^4 \,{\bar \cW}^2 )^2 \Big\} \non \\
&&  -\frac{1}{36}\, {\bar \cD}^4\, (\cW^3 \,\Box {\bar \cW}^3 ) 
~~+~~\cO(\cW^8)~,
\label{cY}
\eea
while $\cS_{\rm int}$ reads
\bea
\cS_{\rm int} &=& 
{ 1 \over 8} \,  \int {\rm d}^{12} \cZ \, 
\cW^2\,{\bar \cW}^2\, \Bigg\{ 1 + 
\hf \, \Big( \cD^4 \cW^2 + {\bar \cD}^4 {\bar \cW}^2 \Big)\label{5-int}   \\
&+&\frac{1}{4} \, 
\Big( (\cD^4 \cW^2)^2 + ({\bar \cD}^4 {\bar \cW}^2)^2 \Big)
+ \frac{3}{4}\, (\cD^4 \cW^2)({\bar \cD}^4 {\bar \cW}^2)
\Bigg\} \non \\
&+&{ 1 \over 24} \,  \int {\rm d}^{12} \cZ \, 
\Bigg\{ {1 \over 3} \cW^3 \Box {\bar \cW}^3 
+\hf (\cW^3 \Box {\bar \cW}^3) {\bar \cD}^4 {\bar \cW}^2 
+\hf ({\bar \cW}^3 \Box \cW^3) \cD^4 \cW^2 
+{ 1 \over 48} \cW^4 \Box^2 {\bar \cW}^4 \Bigg\} \non \\
&+&~~\cO(\cW^{10})~. \non
\eea 
The expression in the first two lines of (\ref{5-int})
comes from the  expansion of (\ref{n=2bi})
in powers of $\cW$ and its conjugate to the order indicated. 
As concerns the expression in the third line of (\ref{5-int}),
it is not present in the power series expansion of (\ref{n=2bi}), 
but it is required for invariance under transformations
(\ref{shifts2}). It is also worth noting that
the expression in the first line of (\ref{cY}) coincides
with the decomposition of $(-\cX )$ (\ref{n=2con}) 
to the given order.

Our conclusion is that 
the system (\ref{n=2bi}), (\ref{n=2con})
cannot be identified with the correct 
$\cN=2$ supersymmetric D3-brane world-volume 
action, and the problem
of constructing such an action is still open.

A natural possibility to look for
$\cN=2$ supersymmetric BI action, advocated in \cite{Ket2},
is first to derive a manifestly $(1,0)$ supersymmetric
BI action in six dimensions and, then to dimensionally reduce
to four dimensions. 
By construction, the resulting four-dimensional  
model should be manifestly $\cN=2$ supersymmetric and
invariant under constant shift transformations 
$\cW \to \cW + \s$, without any nonlinear terms.
However, the problem of constructing
the manifestly $(1,0)$ supersymmetric
BI action in six dimensions is not simple.
In $d=6$ there exists an off-shell formulation 
for the $(1,0)$ vector multiplet \cite{HST}.
But super-extensions of $F^2, ~F^4$ and $F^6$
terms, which appear in the decomposition of the
$d=6$ BI action, cannot be represented by 
integrals over $(1,0)$ superspace or its subspace.
The super-extension of $F^2$ term was already derived in
\cite{HST}. As to the super-extensions of $F^4$ and $F^6$
terms, candidates were proposed in \cite{Ket2}.
Unfortunately, the proof of their invariance 
under $(1,0)$ supersymmetry transformations was 
based on the use of the identity (here we follow
the $d=6$ notation of \cite{HST})
$D_{\a \,(i} \left\{ W^{[\b}_j W^{\g ]}_{k)} \right\} = 0$, 
which holds {\it on-shell} \cite{HST}, and not {\it off-shell}
as claimed in \cite{Ket2}.
Therefore, the super-extensions of $F^4$ and $F^6$ terms
proposed in \cite{Ket2} are not invariant 
under $(1,0)$ supersymmetry transformations.
Thus the problem of constructing a   
manifestly $(1,0)$ supersymmetric
BI action is six dimensions remains unsolved.
If such an action exists, its dimensional 
reduction to $d=4$ will be manifestly supersymmetric, 
but not all terms in the action can be represented as integrals
over $\cN=2$ superspace or its supersymmetric subspaces.

\vskip.5cm

\noindent
{\bf Acknowledgements} \hfill\break
We are grateful to Evgeny Ivanov, Dima Sorokin
and Arkady Tseytlin for their interest in this project. 
We thank Paolo Aschieri for helpful discussions
on duality rotations. 
Support from DFG-SFB-375,
from GIF, the German-Israeli foundation 
for Scientific Research and from the EEC under TMR
contract ERBFMRX-CT96-0045 is gratefully acknowledged.
This work was also supported in part
by the NATO collaborative research grant PST.CLG 974965,
by the RFBR grant No. 99-02-16617, by the INTAS grant
No. 96-0308 and  by the DFG-RFBR grant No. 99-02-04022.

\setcounter{section}{0}
\setcounter{subsection}{0}

\appendix{Derivation of the self-duality equation}

Eq. (\ref{GZ}) is derived as follows.
For an infinitesimal U(1) duality rotation, we have
\be
\tilde{G}'_{ab}(F') = \tilde{G}_{ab}(F) - \l \tilde{F}_{ab} 
= \tilde{G}_{ab}(F)
+ 2 \frac{\pa}{\pa F^{ab}}\; \left( 
- \frac{1}{4}\l \, F \cdot \tilde{F} \right)~,
\label{var-1}
\ee
where we have used the infinitesimal version of eq.(\ref{U(1)-duality}). 
At the same time, from the definition of 
$\tilde{G}'(F')$ it follows
\be
\tilde{G'} (F') = 2 \, \frac{\pa L(F')}{\pa F'}
= 2\left( \frac{\pa }{\pa F'}\, L(F) 
+ \frac{\pa}{\pa F}\, \d L \right)~,
\label{tilde-G'-def}
\ee
where 
\be 
\d L = L(F') - L(F)~.
\ee
Using $F' = F + \l G$, 
one can express 
$\pa / \pa F'$ on the right-hand side of (\ref{tilde-G'-def})
via $\pa / \pa F$ with the  result 
\be 
\tilde{G}'_{ab}(F') = \tilde{G}_{ab}(F)
+ 2\frac{\pa}{\pa F^{ab}}
\left( \d L - \frac{1}{4} \l \, 
G \cdot \tilde{G} 
\right)~.
\label{var-2}
\ee
Comparing eqs. (\ref{var-1}) and (\ref{var-2}) gives
\be 
\d L = \frac{1}{4} \l \, 
( G \cdot \tilde{G} - F \cdot \tilde{F} )~.
\label{var-3}
\ee 
On the other hand, the Lagrangian can be varied directly to give
\be
\d L = \frac{\pa L}{\pa F^{ab}} \, \d F^{ab} = 
\hf \l\, \tilde{G}\cdot G.
\label{var-4}
\ee
This is consistent with eq. (\ref{var-3}) iff
the self-duality equation (\ref{GZ}) holds.

\end{document}